\documentclass[10pt]{amsart}
\usepackage{amsmath,amsfonts,amscd,amssymb,epsf}
\usepackage[small]{caption}

\newtheorem{theorem}{Theorem}[section]

\theoremstyle{definition}

\theoremstyle{remark} \newtheorem{remark}[theorem]{Remark}
\numberwithin{equation}{section}

\newcommand{\Z}{{\mathbb{Z}}}

\newcommand{\M}{\mathcal{M}}

\newcommand{\C}{\mathbb{C}}

\newcommand{\R}{\mathbb{R}}

\newcommand{\pa}{\partial}

\newcommand{\vep}{\varepsilon}

\begin{document}
\title[Casimir energy in closed rectangular cavities]{Finite temperature
Casimir energy in  closed rectangular cavities: a rigorous
derivation based on zeta function technique}
\author{S.C. Lim$^1$}\author{L.P.
Teo$^2$}
\maketitle

\noindent {\scriptsize \hspace{1cm}$^1$Faculty of Engineering,
Multimedia University, Jalan Multimedia, }

\noindent {\scriptsize \hspace{1.1cm} Cyberjaya, 63100, Selangor
Darul Ehsan, Malaysia.}

\noindent {\scriptsize \hspace{1cm} $^2$Faculty of Information
Technology, Multimedia University, Jalan Multimedia,}

\noindent{\scriptsize \hspace{1.1cm} Cyberjaya, 63100, Selangor
Darul Ehsan, Malaysia.}

\noindent{\scriptsize \hspace{1.1cm} E-mail: $^1$sclim@mmu.edu.my
and $^2$lpteo@mmu.edu.my}

 \begin{abstract}
We  derive rigorously explicit formulas of  the Casimir free energy
at finite temperature for massless scalar field  and electromagnetic
field confined in a  closed rectangular cavity with different
boundary conditions by   zeta regularization method. We study both
the low and high temperature expansions of the free energy. In each
case, we write the free energy as a sum of a polynomial in
temperature plus exponentially decay terms. We show that the free
energy is always a decreasing function of temperature. In the cases
of massless scalar field with Dirichlet boundary condition and
  electromagnetic field, the zero temperature Casimir free energy
might be positive. In each of these cases, there is a unique
transition temperature (as a function of the side lengths of the
cavity) where the Casimir energy change from positive to negative.
When the space dimension is equal to two and three, we show
graphically the dependence of this transition temperature on the
side lengths of the cavity. Finally we also show that   we can
obtain the results for a non-closed rectangular cavity by   letting
the size of some directions of a  closed cavity going to infinity,
and we find that these results agree with the usual integration
prescription adopted by other authors.

\vspace{0.5cm} \noindent PACS numbers: 11.10.Wx
\end{abstract}
\section{Introduction}

 The Casimir effect was predicted   in 1948 \cite{Cas} as an effect due to vacuum fluctuation of quantum
 fields. When attempting to calculate the Casimir energy, one inevitably faces the problem of summing a divergent series.
There have been a number of different regularization methods
proposed and used to regularize the infinite sum to extract a
physical finite quantity. Among these methods, zeta regularization
techniques have been widely used recently. One can see for example
the articles \cite{BV, KZ, ER2, AB, E4,CFT, SS} and the books by
Elizalde et al \cite{E1, ET} and Kirsten \cite{K}. This method has
been extended to calculate the Casimir energy at finite temperature
\cite{Kirsten2, ER, OS, Ch1, NLS}. Historically, Casimir effect was
calculated for electromagnetic field confined between two infinitely
conducting parallel plates in four dimensional space--time. Later
on, Casimir energy has been calculated for scalar field, spin $1/2$
field and electromagnetic field in more  general space--time. Among
the different geometries of space that have been under
consideration,  rectangular cavities of different dimensions are
among the most extensively studied \cite{Ch1, AW, CNS, Zheng2,
Actor2, Actor1, Zheng, Mac, Zh1,  Inui1, Inui2, Ed1, HJV, JVH, 4},
partly due to the simple geometry and also the well-developed
mathematical tools. Various aspects of the effect, such as the low
and high temperature expansions of the Casimir energy or force
\cite{Kirsten2, ER, OS, AW, 12, PMG}, the attractive or repulsive
nature of the Casimir force \cite{Ch1, AW, CNS, Zh1, Inui2, Li}, the
effect of extra dimension \cite{OS, Ch1, AW, Zh1}, etc, have been
discussed.

A $p$-dimensional rectangular cavity inside a $d$-dimensional space
is a space of the form
$\Omega_{p,d}=[0,L_1]\times\ldots\times[0,L_p]\times\R^{d-p}$. When
$d=p$, we say that the cavity is closed, and when $p<d$, the cavity
is non-closed.
 The paper by Ambj{\o}rn and
Wolfram \cite{AW} can be considered as the pioneer work in the
calculation and discussion of Casimir effects at finite temperature
for massless scalar field and electromagnetic field confined within
a rectangular cavity. By using dimensional regularization technique,
they found that the Casimir energy can be expressed using Epstein
zeta function whose analytic continuation is well--known. Ambj{\o}rn
and Wolfram were also able to obtain the high and low temperature
expansions of the free energy by using the Chowla-Selberg formula
\cite{CS} for Epstein zeta function. Their formulas work for $p<d$,
whereas for the case of closed cavity (i.e. the $p=d$ case), they
modified the $p<d$ formulas to remove divergencies based on physical
arguments. However, the divergencies for the high and low
temperature expansions were removed separately and they did not
justify that the two results coincide at any temperature.

Special cases of the results of Ambj{\o}rn and Wolfram have been
reproduced and extended by several authors using zeta regularization
or other methods, see for examples \cite{OS, Ch1, CNS, Zheng2,
Zheng, Zh1, Ed1, HJV, Li}. In particular, there have been an
extensive study of the Chowla-Selberg formula for general Epstein
zeta function \cite{ER2, E4, Kirsten2, ER, 12, 13, E5, E6, E7, E8}
with the aim to obtain the low and high temperature expansions of
the Casimir energy. However, to the best of our knowledge, no one
has derived the Casimir energy for fields confined in closed
rectangular cavities correctly (without divergent terms) purely by
zeta regularization techniques. One can read for example the third
paragraph in the introduction of \cite{JVH}, where they pointed out
this divergency problem in some of the literatures (e.g. \cite{TS}).
In \cite{JVH}, the authors also mentioned that it is desirable to
obtain a closed formula for the free energy of the electromagnetic
field confined in a three dimensional rectangular cavity that is
valid for all temperature.

In this paper, we solve a more general problem. We derive the
Casimir free energy at finite temperature for massless scalar fields
and electromagnetic fields confined in a closed rectangular cavity
with different boundary conditions, by employing zeta regularization
techniques. We derive explicit formulas for the free energy, in the
low and high temperature regions respectively. However, we want to
emphasize that  both the low and high temperature formulas are valid
at all temperature. Their difference lies in the manifestation of
the leading behavior of the free energy at low and high temperature
respectively. The advantage of using the zeta regularization
approach is that we can derive formulas that work for any dimension
$d\geq 2$ at one shot. With the further help rendered by the
Chowla-Selberg formula, we can compute the free energy  effectively.
We show some results graphically when $d=2$ and $d=3$. On the other
hand, we also study some behavior of the free energy using the
formulas we derive. In particular, we find that the free energy is
always a decreasing function of temperature. In the cases of
massless scalar field with periodic and Neumann boundary conditions,
the zero temperature free energy is always negative. Therefore, the
free energy is negative at all temperature. In the cases of massless
scalar field with Dirichlet boundary condition and electromagnetic
fields, the zero temperature free energy can be positive. We study
the cases when $d=2$ and $d=3$, and we leave a more detail study of
the general cases to another paper. When the zero temperature free
energy is positive, we can conclude from the decreasing behavior of
the free energy that there is a unique transition temperature
(depending on the side lengths of the cavity) where the sign of the
free energy changes from positive to negative. We show graphically
the dependence of this transition temperature on the side lengths
when $d=2$ and $d=3$.  In the last section, we  show how to obtain
the corresponding results for a non-closed rectangular cavity
$\Omega_{p,d}$
 by letting the size of $d-p$ directions of a closed cavity going to infinity.
We find that our results are in agreement with those based on the
method of changing the summation in $d-p$ directions to integration,
which is commonly adopted by other authors.

\section{Casimir Energy at finite temperature}\label{sec2}

For a massless scalar field $\phi$ in $d$-dimensional space $\Omega$
maintained in thermal equilibrium at temperature $T$, the Helmholtz
free energy is conventionally defined as
\begin{align*}
F=-\frac{1}{\beta}\log Z,
\end{align*}where $\beta=1/T$ and $Z$ is the partition function
given by
\begin{align}\label{eq1}
Z=\prod_{\mathbf{k}}\!'\frac{e^{-\beta\omega_{\mathbf{k}}/2}}{1-e^{-\beta\omega_{\mathbf{k}}}}.
\end{align}Here $\omega_{\mathbf{k}}$ is the frequency associated with
the eigenmode $\phi_{\mathbf{k}}$ of the field, and the symbol
$\,'\,$ in the product means that the term $\omega_{\mathbf{k}}=0$
is to be omitted. More precisely, the free energy $F$ is equal to
\begin{align}\label{eq2}
F=-\frac{1}{\beta}\log
Z=\frac{1}{2}\sum_{\mathbf{k}}\!'\omega_{\mathbf{k}}+\frac{1}{\beta}\sum_{\mathbf{k}}\!'\log\left(1-e^{-\beta\omega_{\mathbf{k}}}\right).
\end{align}The first term
$$F^0=\frac{1}{2}\sum_{\mathbf{k}}\!'\omega_{\mathbf{k}}$$ is the zero
temperature contribution to the free energy, also known as Casimir
free energy. The summation is divergent and regularization is needed
to obtain a finite value. There are various regularization
techniques that have been employed. One of the conventional methods
is to introduce the zeta function $\zeta_{\Omega}(s)$ (see e.g.
\cite{E1, K}):
\begin{align*}
\zeta_{\Omega}(s) =\sum_{\mathbf{k}}\!'\omega_{\mathbf{k}}^{-2s},
\hspace{1cm}\text{Re}\, s> \frac{d}{2}.
\end{align*}It is well known that $\zeta_{\Omega}(s)$ can be
analytically continued to the complex plane with possible simple
poles at $s=\frac{d-l}{2}$, $l=0,1,2\ldots$. In the case
$\zeta_{\Omega}(s)$ is regular at $s=-1/2$, we can define
\begin{align}\label{eq3}
F^0=\frac{1}{2}\sum_{\mathbf{k}}\!'\omega_{\mathbf{k}}=\frac{1}{2}\zeta_{\Omega}\left(-\frac{1}{2}\right).
\end{align} In general, as was proposed by Blau and Visser \cite{BV},
one should introduce a constant $\lambda$ with dimension
(length)$^{-1}$ and define
\begin{align*}
F^0=&\frac{1}{2}\text{P.P.}_{s\rightarrow
-\frac{1}{2}}E_{\Omega}(s)\\
=&\frac{1}{4}\left(\lim_{\vep\rightarrow
0}E_{\Omega}\left(-\frac{1}{2}+\vep\right)+E_{\Omega}\left(-\frac{1}{2}-\vep\right)\right),
\end{align*}
where P.P. means principal part and $E_{\Omega}(s)$ is the
normalized zeta function
\begin{align*}
E_{\Omega}(s)=\lambda\sum_{\mathbf{k}}\!'\left(\frac{\omega_{\mathbf{k}}^2}{\lambda^2}\right)^{-s}=\lambda^{1+2s}
\zeta_{\Omega}(s).
\end{align*}Since $\zeta_{\Omega}(s)$ may
have a simple pole at $s=-1/2$, we can write
\begin{align}\label{eq4}
\zeta_{\Omega}(s)=\frac{r_1}{s+\frac{1}{2}}+ r_0+
O\left(s+\frac{1}{2}\right).
\end{align}A straightforward computation gives
\begin{align*}
F^0=&\frac{1}{2}\left(r_0+r_1\log\lambda^2\right).
\end{align*}If $\zeta_{\Omega}(s)$ is regular at $s=-1/2$, $r_1=0$,
$r_0=\zeta_{\Omega}(-1/2)$ and we get back the definition
\eqref{eq3}.

 The second term in \eqref{eq2}
$$\Delta F = \frac{1}{\beta}\sum_{\mathbf{k}}\!'\log\left(1-e^{-\beta\omega_{\mathbf{k}}}\right),$$
 is known as the thermal correction to the free energy. Due to
the exponential term, it is a finite sum. Hence if we are interested
in the low temperature behavior of the free energy, we can use the
expression
\begin{align}\label{eq5}
F=-\frac{1}{\beta}\log
Z=\frac{1}{2}\left(r_0+r_1\log\lambda^2\right)+
\frac{1}{\beta}\sum_{\mathbf{k}}\!'\log\left(1-e^{-\beta\omega_{\mathbf{k}}}\right).
\end{align}However, this expression is not convenient for studying
the high temperature behavior of the free energy.

\begin{remark}\label{re1}
Differentiate \eqref{eq5} with respect to $\beta$, we find that
\begin{align*}
\frac{\pa F}{\pa
\beta}=-\frac{1}{\beta^2}\sum_{\mathbf{k}}\!'\log\left(1-e^{-\beta\omega_{\mathbf{k}}}\right)
+\frac{1}{\beta}\sum_{\mathbf{k}}\!'
\frac{\omega_{\mathbf{k}}e^{-\beta\omega_{\mathbf{k}}}}{1-e^{-\beta\omega_{\mathbf{k}}}}\geq
0.
\end{align*}Therefore the free energy is always an increasing
function of $\beta$, and thus a decreasing function of the
temperature $T$. Hence, if the zero temperature free energy $F^0$ is
negative, then the free energy $F$ will be negative for all
temperature.
\end{remark}

It has been  taken for granted (or taken as definition) that the
partition function $Z$ can be calculated using the path integral
\begin{align}\label{eq6}
\mathcal{Z}=\int\limits_{ \substack{\text{boundary}\\\text{
conditions} } } D\varphi \exp\left(-\int_0^{\beta}\int_{\Omega}
\varphi(\mathbf{x},t)(-\square_E)\varphi(\mathbf{x},t)
d^{d}\mathbf{x}dt\right)=\det
\left(-\frac{1}{\mu^2}\square_E\right)^{-1/2},
\end{align}where $$\square_E=\frac{\pa^2}{\pa t^2}+\sum_{i=1}^{d}\frac{\pa^2}{\pa
x_i^2}$$ is the $(d+1)$-dimensional Euclidean d-Alembertian operator
and $\mu$ is a normalization constant with the dimension of mass. In
the imaginary time formalism (or Matsubara formalism) of finite
temperature field theory, one imposes periodic boundary condition
with period $\beta$ in time direction. In the spatial direction,
$\varphi$ is assumed to have the same boundary condition as $\phi$.
The eigenvalues of $-\square_E$ are then given by
\begin{align*} \Lambda_{n,\mathbf{k}}=\left(\frac{2\pi
n}{\beta}\right)^2+\omega_{\mathbf{k}}^2, \hspace{1cm} n\in\Z.
\end{align*}
Using the zeta regularization method, one defines
\begin{align}\label{eq7}
\zeta(s)=\sum_{n=-\infty}^{\infty}\sum_{\mathbf{k}}\!'\left(\left(\frac{2\pi
n}{\beta}\right)^2+\omega_{\mathbf{k}}^2\right)^{-s}
\end{align}which is an analytic function of $s$ when $\text{Re}\;s>(d+1)/2$.
Here the symbol  $\,'\,$ in the double summation means that a term
where $(n, \omega_{k})=(0,0)$ should be omitted. One then
analytically continue $\zeta(s)$ to the complex plane and the
logarithm of \eqref{eq6} is then equal to
\begin{align}\label{eq8}
\log\mathcal{Z}=
\frac{1}{2}\zeta'(0)+\frac{1}{2}\left(\log\mu^2\right)\zeta(0).
\end{align}Most people set $\mu^2=1$ and claim that $\log
Z=\log\mathcal{Z}$. However, we are going to show that this is not
true when there are some modes $\phi_{\mathbf{k}}$ with
$\omega_{\mathbf{k}}=0$. Since in the definition of the partition
function \eqref{eq1}, we omit the terms where
$\omega_{\mathbf{k}}=0$, therefore it is natural to single out the
contribution from $\omega_{\mathbf{k}}=0$ terms and write
\eqref{eq7} as
\begin{align*}
\zeta(s)=
2\mathcal{N}\left(\frac{2\pi}{\beta}\right)^{-2s}\zeta_R(2s)+\hat{\zeta}(s),
\end{align*}where $\zeta_R(s)=\sum_{n=1}^{\infty}n^{-s}$ is the Riemann zeta function,
$\mathcal{N}$ is the number of modes $\phi_{\mathbf{k}}$ with $\omega_{\mathbf{k}}=0$ and\begin{align*}
\hat{\zeta}(s)=\sum_{n=-\infty}^{\infty}\sum_{\omega_{\mathbf{k}}\neq
0}\left(\left(\frac{2\pi
n}{\beta}\right)^2+\omega_{\mathbf{k}}^2\right)^{-s}.
\end{align*}
It is well known that $\zeta_R(s)$ has analytic continuation to the
whole complex plane with single pole at $s=1$. On the other hand,
using standard techniques, for $\text{Re}\; s > (d+1)/2$,
$\hat{\zeta}(s)$ is analytic and is given explicitly by
\begin{align*}
\hat{\zeta}(s)=&\frac{\beta\Gamma\left(s-\frac{1}{2}\right)}{2\sqrt{\pi}\Gamma(s)}\zeta_{\Omega}\left(s-\frac{1}{2}\right)
+\frac{2\beta}{\sqrt{\pi}\Gamma(s)}\sum_{n=1}^{\infty}\sum_{\omega_{\mathbf{k}}\neq
0} \left(\frac{\beta
n}{2\omega_{\mathbf{k}}}\right)^{s-1/2}K_{s-1/2}\left(\beta n
\omega_{\mathbf{k}}\right).
\end{align*}Here $K_{\nu}(z)$ is the modified Bessel function of second kind (see e.g., \textbf{3.471} in \cite{GR}).
From this, we find that \eqref{eq8} is given
by\begin{align}\label{eq10}
\log\mathcal{Z}=-\frac{\mathcal{N}}{2}\log(\beta\mu)^2-\beta\Biggl(
\frac{r_1}{2}\log\mu^2 +(1-\log 2) r_1
+\frac{r_0}{2}+\frac{1}{\beta}\sum_{\omega_{\mathbf{k}}\neq 0}
\log\left(1-e^{-\beta\omega_{\mathbf{k}}}\right)\Biggr),
\end{align}whereas
\begin{align}\label{eq11}
\log\hat{\mathcal{Z}}:=&\frac{1}{2}\hat{\zeta}'(0)+\frac{1}{2}\left(\log\mu^2\right)\hat{\zeta}(0)\\
=&-\beta\Biggl( \frac{r_1}{2}\log\mu^2 +(1-\log 2) r_1
+\frac{r_0}{2}+\frac{1}{\beta}\sum_{\omega_{\mathbf{k}}\neq 0}
\log\left(1-e^{-\beta\omega_{\mathbf{k}}}\right)\Biggr).\nonumber
\end{align}Compare these expressions with \eqref{eq5}, we note
that when $\mathcal{N}\neq 0$ (i.e. in the presence of
$\omega_{\mathbf{k}}=0$ modes), $\log Z\neq \log\mathcal{Z}  $, but
$\log Z =\log\hat{\mathcal{Z}}$ if we identify $\lambda$ with
$e\mu/2$.

It has been noticed by several authors (see e.g. \cite{BV, AB, OS})
that the Casimir energy at zero temperature can be defined by
$$\lim_{\beta\rightarrow
\infty}\left(-\frac{1}{\beta}\log\mathcal{Z}\right).$$ In view of
what we have obtained  above, due care has to be taken in the
presence of $\omega_{\mathbf{k}}=0$ modes. In this case, we should
replace $\log\mathcal{Z}$ by $\log\hat{\mathcal{Z}}$. From
\eqref{eq5}, \eqref{eq10} and \eqref{eq11}, we can write
\begin{align*}
F=-\frac{1}{\beta}\log Z =-\frac{1}{\beta}\left(\log
\mathcal{Z}+\frac{\mathcal{N}}{2}\log\left(\beta\mu\right)^2\right)
\end{align*}and therefore\begin{align*}
F^0=\lim_{\beta\rightarrow
\infty}\left(-\frac{1}{\beta}\left[\log\mathcal{Z}+\frac{\mathcal{N}}{2}\log\left(
\beta\mu\right)^2 \right]\right)
\end{align*}with the identification $\lambda=e\mu/2$.

The constants $\mu$ or $\lambda$ contribute ambiguities to the
Casimir free energy. However, in most of the cases of interest, the
function $\zeta_{\Omega}(s)$ is regular at $s=-1/2$. This is
equivalent to $r_1=0$. Using the zeta function $\zeta(s)$, we can
characterize such cases by $\zeta(0)=-\mathcal{N}$. Hence if
$\zeta(0)=-\mathcal{N}$, the Casimir energy turns out to be
independent of $\mu$ or $\lambda$ and can be calculated by using
\begin{align}\label{eq12}
F=-\frac{1}{2\beta}\left(\zeta'(0)+2\mathcal{N}\log\beta\right),
\end{align}in contrast to the usual prescription
$F=-\frac{1}{2\beta}\zeta'(0)$.

The expression for $\log\mathcal{Z}$ \eqref{eq10}, with the presence
of $\omega_{\mathbf{k}}=0$ terms has been obtained in \cite{OS}.
However, in \cite{OS}, the discrepancy between $\mathcal{Z}$ and the
thermodynamic partition function $Z$ was not emphasized. On the
other hand, a computation similar to what we perform above was done
in \cite{KZ}, without taking into consideration the
$\omega_{\mathbf{k}}=0$ terms.

In some of the studies (e.g. \cite{Ch1}), the (internal) energy $E$
of the system was calculated instead of the free energy $F$. They
are related by
\begin{align}\label{eq17_1}
E=-\frac{\pa(\beta F)}{\pa\beta}.
\end{align}Another important thermodynamic quantity--the entropy
$S$, can be calculated from the free energy by the formula
\begin{align}\label{eq17_2}
S=-\frac{\pa F}{\pa T} =\beta^2\frac{\pa F}{\pa\beta}.
\end{align}In view of Remark \ref{re1}, it is always non-negative. In the following, we will only compute the free energy
explicitly. We leave the readers to work out the  energy and entropy
themselves by using these two formulas.

\begin{remark}
For the sake of convenience of presentation, in this section, we
have assumed that $\phi$ is a massless scalar field. However, the
same reasoning works for other quantum fields.
\end{remark}
\section{Homogeneous Epstein Zeta function}

Now we want to  compute the derivative at zero of the Epstein zeta
function using Chowla-Selberg formula. In association with the
application of zeta regularization method, Chowla-Selberg formula
has been extensively used to  express Epstein zeta function in the
form which facilitates the study of the function in certain limits
\cite{ER2, E4, Kirsten2, ER, 12, 13, E5, E6, E7, E8}. However, we
are unaware of anything done  regarding the explicit computation of
the derivative at zero of the homogeneous Epstein zeta function.

In this paper, we only consider the homogeneous Epstein zeta
function in $n$ variables in the following form:
\begin{align*} Z_{E,n}(s; a_1,\ldots, a_n)= \sum_{(k_1,\ldots,
k_n)\in\Z^{n}\setminus\{ \mathbf{0}\}} \frac{1}{([a_1k_1]^2+\ldots
+[a_nk_n]^2)^s}.
\end{align*}
This sum is convergent for $s>n/2$. Under a scaling $a_i\mapsto
\lambda a_i$, we have
\begin{align}\label{eq30_1}
Z_{E,n}(s; \lambda a_1,\ldots, \lambda a_n)=\lambda^{-2s}Z_{E,n}(s;
a_1,\ldots, a_n).
\end{align}

To find the derivative at $s=0$, we first derive the Chowla-Selberg
formula for Epstein zeta function. For fixed $1\leq m \leq n-1$, we
can write
\begin{align*}
Z_{E,n}(s;a_1,\ldots, a_n)=&Z_{E,m}(s; a_1, \ldots,
a_m)\\&+\sum_{(k_1, \ldots, k_m)\in \Z^m}\sum_{(k_{m+1}, \ldots,
k_n)\in\Z^{n-m}\setminus\{\mathbf{0}\}}\frac{1}{([a_1k_1]^2+\ldots
+[a_n k_n]^2)^s}.
\end{align*}For the second term, we have
\begin{align}\label{eq4_6}
&\sum_{\mathbf{k}\in\Z^m\times(\Z^{n-m}\setminus\{\mathbf{0}\})}\frac{1}{([a_1k_1]^2+\ldots
+[a_nk_n]^2)^s}\\
=&\frac{1}{\Gamma(s)}\int_0^{\infty}
t^{s-1}\sum_{\mathbf{k}\in\Z^m\times(\Z^{n-m}\setminus\{\mathbf{0}\})}e^{-t([a_1k_1]^2+\ldots
+[a_mk_m]^2)}e^{-t([a_{m+1}k_{m+1}]^2+\ldots
+[a_nk_n]^2)}dt\nonumber\\=&\frac{\sqrt{\pi}^m}{\left[\prod_{j=1}^m
a_j\right]\Gamma(s)}\int_0^{\infty}
t^{s-\frac{m}{2}-1}\sum_{\mathbf{k}\in\Z^m\times(\Z^{n-m}\setminus\{\mathbf{0}\})}
e^{-\frac{\pi^2}{t}\sum_{j=1}^m\left[\frac{k_j}{a_j}\right]^2-t\sum_{l=m+1}^n
[a_lk_l]^2}\nonumber\\
=&\frac{\pi^{m/2}\Gamma\left(s-\frac{m}{2}\right)}{\left[\prod_{j=1}^ma_j\right]\Gamma(s)}Z_{E,
n-m}\left(s-\frac{m}{2}; a_{m+1}, \ldots,
a_n\right)+\frac{1}{\Gamma(s)}T_{n,m}(s; a_1, \ldots, a_n),\nonumber
\end{align}where\begin{align*}
&T_{n,m}(s;a_1,\ldots,
a_n)=\frac{2\pi^{s}}{\left[\prod_{j=1}^ma_j\right]}\sum_{\mathbf{k}\in(\Z^m\setminus
\{\mathbf{0}\})\times(\Z^{n-m}\setminus\{\mathbf{0}\})}\\&\left(\frac{\sum_{j=1}^m\left[\frac{k_j}{a_j}\right]^2}
{\sum_{l=m+1}^n[a_{l}k_{l}]^2}\right)^{\frac{2s-m}{4}}K_{s-\frac{m}{2}}
\left(2\pi\sqrt{\left(\sum_{j=1}^{m}\left[
\frac{k_j}{a_j}\right]^2\right)\left(\sum_{l=m+1}^n[a_{l}k_{l}]^2\right)}\right).
\end{align*}Combine together, we have the Chowla-Selberg formula
\begin{align}\label{eq31}
&Z_{E,n}(s;a_1,\ldots, a_n)=Z_{E,m}(s; a_1, \ldots,
a_m)\\&\hspace{1cm}+\frac{\pi^{m/2}\Gamma\left(s-\frac{m}{2}\right)}{\left[\prod_{j=1}^ma_j\right]\Gamma(s)}Z_{E,
n-m}\left(s-\frac{m}{2}; a_{m+1}, \ldots,
a_n\right)+\frac{1}{\Gamma(s)}T_{n,m}(s; a_1, \ldots, a_n).\nonumber
\end{align}The function $T_{n,m}(s;
a_1, \ldots, a_n)$ is an analytic function of $s$ on $\C$. Using the
fact that the Riemann zeta function $\zeta_R(s)$ is meromorphic on
$\C$ with a single pole at $s=1$ and the fact that $Z_{E,1}(s;a)=
2a^{-2s}\zeta_R(2s)$, we obtain by recursion a meromorphic extension
of $Z_{E,n}(s; a_1,\ldots, a_n)$ to $\C$ with single pole at
$s=n/2$. On the other hand, one can also use the Chowla-Selberg
formula \eqref{eq31} to prove the reflection formula
\begin{align}\label{eq32}
\pi^{-s}\Gamma(s)Z_{E,n}(s; a_1, \ldots , a_n)
=\frac{\pi^{s-\frac{n}{2}}}{\left[\prod_{j=1}^na_j\right]}\Gamma\left(\frac{n}{2}-s\right)Z_{E,n}\left(\frac{n}{2}-s;
\frac{1}{a_1},\ldots, \frac{1}{a_n}\right)
\end{align}by induction (see e.g.
\cite{AT}).
 Putting $m=1$ and $s=0$ in \eqref{eq31}, using the
reflection formula \eqref{eq32}, the fact that
$(1/\Gamma(s))|_{s=0}=(s/\Gamma(s+1))|_{s=0}=0$ and
$\zeta_R(0)=-1/2$, we find that
\begin{align}\label{eq33}
Z_{E, n}(0; a_1, \ldots, a_n)=Z_{E,1}(0;a)=2\zeta_R(0)=-1.
\end{align}On the other hand, if $a_1\leq \ldots\leq  a_n$, by putting $m=n-1$ in the
Chowla-Selberg formula \eqref{eq31}, we obtain by recursion
\begin{align*}
&Z_{E,n}(s; a_1,\ldots, a_n)\\
=&2a_1^{-2s}\zeta_R(2s)
+\frac{2}{\Gamma(s)}\sum_{j=1}^{n-1}\frac{\pi^\frac{j}{2}\Gamma\left(s-\frac{j}{2}\right)}
{a_{j+1}^{2s-j}\prod_{l=1}^{j}a_l} \zeta_R(2s-j)
+\frac{4\pi^s}{\Gamma(s)}\sum_{j=1}^{n-1}
\frac{1}{\prod_{l=1}^{j}a_l}\times\\&\sum_{\mathbf{k}\in
\Z^j\setminus\{\mathbf{0}\}}\sum_{m=1}^{\infty}\frac{1}{(ma_{j+1})^{s-\frac{j}{2}}}\left(\sum_{l=1}^j
\left[\frac{k_l}{a_l}\right]^2\right)^{\frac{s}{2}-\frac{j}{4}}K_{s-\frac{j}{2}}\left(
2\pi m a_{j+1}\sqrt{\sum_{l=1}^j
\left[\frac{k_l}{a_l}\right]^2}\right),
\end{align*}which express the Epstein Zeta function as a sum of
Riemann zeta functions plus a remainder which is a multi-dimensional
series that converges rapidly. This formula can be used to
effectively compute the Epstein zeta function to any degree of
accuracy.

To compute the derivative $Z_{E,n}'(s;a_1,\ldots,a_n)$ at $s=0$, we
differentiate the Chowla-Selberg formula \eqref{eq31} with respect
to $s$ and setting $s=0$. This gives
\begin{align}\label{eq25_10}
&Z_{E,n}'(0;a_1,\ldots, a_n)\\=\nonumber&Z_{E,m}'(0; a_1, \ldots,
a_m)+\frac{\pi^{m/2}\Gamma\left(-\frac{m}{2}\right)}{\left[\prod_{j=1}^ma_j\right]}Z_{E,
n-m}\left(-\frac{m}{2}; a_{m+1}, \ldots, a_n\right)+R_{n,m}( a_1,
\ldots, a_n)\\=\nonumber&Z_{E,m}'(0; a_1, \ldots,
a_m)+\frac{\pi^{-n/2}\Gamma\left(\frac{n}{2}\right)}{\left[\prod_{j=1}^na_j\right]}Z_{E,
n-m}\left(\frac{n}{2}; \frac{1}{a_{m+1}}, \ldots,
\frac{1}{a_n}\right)+R_{n,m}( a_1, \ldots, a_n),
\end{align}where
\begin{align}\label{eq30_4}
&R_{n,m}( a_1, \ldots,
a_n)=\frac{2}{\left[\prod_{j=1}^ma_j\right]}\sum_{\mathbf{k}\in(\Z^m\setminus
\{\mathbf{0}\})\times(\Z^{n-m}\setminus\{\mathbf{0}\})}\\&\left(\frac{\sum_{j=1}^m\left[\frac{k_j}{a_j}\right]^2}
{\sum_{l=m+1}^n[a_{l}k_{l}]^2}\right)^{-\frac{m}{4}}K_{\frac{m}{2}}\left(2\pi\sqrt{\left(
\sum_{j=1}^m\left[\frac{k_j}{a_j}\right]^2
\right)\left(\sum_{l=m+1}^n[a_{l}k_{l}]^2\right)}\right).\nonumber
\end{align}
Using \eqref{eq30_1} and \eqref{eq33}, we find that under the
scaling $a_i\mapsto \lambda a_i$, we have
\begin{align}\label{eq30_2}
Z_{E,n}'(0; \lambda a_1,\ldots, \lambda
a_n)=2\log\lambda+Z_{E,n}'(0; a_1,\ldots, a_n).
\end{align}

\section{Massless scalar field inside closed rectangular cavity}
In this section,  Casimir energy at finite temperature for massless
scalar field confined within a closed rectangular cavity of
dimension $d\geq 2$ will be derived. Using the notations in Section
\ref{sec2}, the $d$-dimensional space $\Omega$ is the rectangular
box $[0, L_1]\times\ldots [0, L_{d}]$ with volume $V=L_1\ldots L_d$.
Without loss of generality, we assume that $0<L_1\leq \ldots \leq
L_d$. We are going to consider the following boundary conditions for
the field $\phi$: A) Periodic boundary condition, B) Dirichlet
boundary condition, C) Neumann boundary condition.

\vspace{0.5cm}\noindent \textbf{A) Periodic Boundary Condition.}~
Consider the periodic boundary condition with $\phi(x_1, \ldots,
x_j+L_j, \ldots, x_d)= \phi(x_1,\ldots, x_j, \ldots, x_d)$ for all
$1\leq j\leq d$. In this case, the eigenmodes of $\phi$ are
\begin{align*}
\phi_{\mathbf{k}}(\mathbf{x})= e^{ i\left(\frac{2\pi
k_1x_1}{L_1}+\ldots +\frac{2\pi  k_d x_d}{L_d}\right)}, \hspace{1cm}
\mathbf{k}\in \Z^d.
\end{align*}The corresponding zeta function $\zeta(s)$ is
\begin{align*}
\zeta_{P, d}(s; L_1,\ldots,L_d) =&\sum_{(m,\mathbf{k})\in
\Z^{d+1}\setminus\{\mathbf{0}\}}\left(\left(\frac{2\pi
m}{\beta}\right)^2+\left(\frac{2\pi k_1}{L_1}\right)^2+\ldots
+\left(\frac{2\pi k_d }{L_d}\right)^2\right)^{-s}\\
=&Z_{E, d+1}\left(s; \frac{2\pi}{\beta}, \frac{2\pi}{L_1}, \ldots,
\frac{2\pi}{L_d}\right)
\end{align*}and there is $\mathcal{N}=1$ zero modes of $\phi$
corresponding to $\mathbf{k}=\mathbf{0}$. By \eqref{eq33},
$\zeta_{P,d}(0 ; L_1,\ldots,L_d)=-1=-\mathcal{N}$. Therefore by
\eqref{eq12}, the Casimir free energy is given by
\begin{align}\label{eq30_3}
F_{P}(L_1, \ldots, L_d)= -\frac{1}{2\beta}Z_{E, d+1}'\left(0;
\frac{2\pi}{\beta }, \frac{2\pi}{L_1 }, \ldots,
\frac{2\pi}{L_d}\right)-\frac{1}{\beta}\log\beta.
\end{align}Using \eqref{eq30_2}, we find that under the simultaneous
scaling $\beta\mapsto \lambda \beta$, $L_i\mapsto \lambda L_i$, the
free energy $F_{P}(L_1, \ldots, L_d)$ transform as
\begin{align}\label{eq17_3}
F_{P}(L_1, \ldots, L_d)\mapsto F_{P}(\lambda L_1, \ldots, \lambda
L_d)=\frac{1}{\lambda}F_{P}(L_1, \ldots, L_d).
\end{align}Therefore, when studying the free energy, we can  define the scaled variables
\begin{align*}
\xi=\frac{\beta}{V^{1/d}}, \hspace{1cm} l_i=\frac{L_i}{V^{1/d}},
\hspace{0.5cm} 1\leq i\leq d,
\end{align*}called the scaled temperature and the scaled side lengths of the cavity respectively.
The function $V^{1/d}F_P(L_1,\ldots, L_d)$ is then a function of
these scaled variables:   \begin{align*} V^{1/d}F_P(L_1,\ldots,
L_d)=-\frac{1}{2\xi}Z_{E, d+1}'\left(0; \frac{2\pi}{\xi },
\frac{2\pi}{l_1 }, \ldots,
\frac{2\pi}{l_d}\right)-\frac{1}{\xi}\log\xi.
\end{align*}with $l_1\ldots l_d=1$.

The Casimir force on the walls $x_j=0$ and $x_{j}=L_j$ is given by
\begin{align}\label{eq5_23_2}
\mathfrak{F}_{j}=-\frac{\pa F}{ \pa L_j}
\end{align}and the corresponding pressure is
\begin{align}\label{eq5_23_3}
P_{j}=\frac{L_j\mathfrak{F}_{j}}{V}.
\end{align}

\vspace{0.2cm}\noindent \textbf{A.1. Low temperature expansion.} ~
By putting $m=1$, $a_1=2\pi/\beta$, $a_j =2\pi/L_{j-1}$ when $2\leq
j\leq d+1$ in \eqref{eq25_10}, we obtain the low temperature
($T=1/\beta\ll 1$) expansion
\begin{align}\label{eq56}
F_{P}(L_1, \ldots, L_d)=& -\frac{L_1\ldots
L_d}{2\pi^{\frac{d+1}{2}}} \Gamma\left(\frac{d+1}{2}\right) Z_{E,d}
\left(\frac{d+1}{2}; L_1,\ldots,
L_d\right)\\&+\frac{1}{\beta}\sum_{\mathbf{k}\in\Z^d\setminus\{\mathbf{0}\}}
\log\left(1-e^{-\beta\sqrt{\left(\frac{2\pi
k_1}{L_1}\right)^2+\ldots+\left(\frac{2\pi
k_d}{L_d}\right)^2}}\right),\nonumber
\end{align}which have the form of \eqref{eq5}. We find directly that the zero temperature Casimir energy is
\begin{align}\label{eq55}
F_{P}^0(L_1, \ldots, L_d)=& -\frac{L_1\ldots
L_d}{2\pi^{\frac{d+1}{2}}} \Gamma\left(\frac{d+1}{2}\right) Z_{E,d}
\left(\frac{d+1}{2}; L_1,\ldots, L_d\right),
\end{align}which agrees with (3.4) in \cite{AW}. A similar result was obtained by Edery  \cite{Ed1}
using multidimensional cut-off technique. By the definition of the
Epstein zeta function, the term \eqref{eq55} is strictly negative.
Remark \ref{re1} then implies the Casimir free energy is then always
negative for all temperature. On the other hand,
 we can compute an explicit upper bound for the thermal correction
 term:
\begin{align*}
\left|\Delta F_{P}(L_1, \ldots,
L_d)\right|=\left|\frac{1}{\beta}\sum_{\mathbf{k}\in\Z^d\setminus\{\mathbf{0}\}}
\log\left(1-e^{-\beta\sqrt{\left(\frac{2\pi
k_1}{L_1}\right)^2+\ldots+\left(\frac{2\pi
k_d}{L_d}\right)^2}}\right)\right|\leq \frac{1}{\beta}\frac{2^d
de^{-\frac{2\pi\beta }{\sqrt{d}L_d}}}{\left(1-e^{-\frac{2\pi\beta
}{\sqrt{d}L_d}}\right)^{d+1}},
\end{align*}which is an exponentially decay term as
$\beta\rightarrow\infty$.

From \eqref{eq30_1} and \eqref{eq55} (or by \eqref{eq17_3}), we see
that under the space scaling $L_i\mapsto\lambda L_i , 1\leq
\lambda\leq d$, the zero temperature free energy transforms as
\begin{align}\label{eq10_3}
F_{P}^0(L_1, \ldots, L_d)\mapsto F_{P}^0(\lambda L_1, \ldots,
\lambda L_d)=\lambda^{-1}F_{P}^0(L_1, \ldots, L_d).
\end{align}Namely, the zero temperature free energy  is inversely
proportional to the dimension of space. This scaling property breaks
down at positive temperature. However, \eqref{eq17_3} shows that
this scaling behavior will hold if  the temperature is also scaled
inversely. On the other hand, differentiating the equation on the
right hand side of \eqref{eq10_3} with respect to $\lambda$ and
setting $\lambda=1$, we get
\begin{align*}
L_1\frac{\pa F^0}{\pa L_1}+\ldots+L_d\frac{\pa F^0}{\pa L_d}=-F^0.
\end{align*}From the definition of pressure
\eqref{eq5_23_3}, we find that at zero temperature, the equation of
state
\begin{align}\label{eq5_23_4}
F^0=(P_1 +\ldots+P_d )V
\end{align}holds. When the cavity is a hypercube (i.e. when
$L_1=\ldots=L_d$), this implies that the zero temperature free
energy $F^0$ always has the same sign as the force and pressure. At
finite temperature, as a correction to \eqref{eq5_23_4},
\eqref{eq17_3} gives us the well-known thermodynamic relation
\begin{align}\label{eq5_23_7}
F=-L_1\frac{\pa F}{\pa L_1}-\ldots-L_d\frac{\pa F}{\pa
L_d}-\beta\frac{\pa F}{\pa \beta}=(P_1+\ldots+P_d)V-TS,
\end{align}where $S$ is the entropy \eqref{eq17_2}.

Using Arithmetic-Geometric inequality, we find that when
$V=L_1\ldots L_d$ is fixed,
\begin{align*}
(L_1k_1)^2+\ldots +(L_d k_d)^2\geq d (k_1\ldots
k_d)^{\frac{2}{d}}V^{\frac{2}{d}},\\
\left(\frac{k_1}{L_1}\right)^2+\ldots\left(\frac{k_d}{L_d}\right)^2\geq
d (k_1\ldots k_d)^{\frac{2}{d}}V^{-\frac{2}{d}},
\end{align*}and equalities hold if and only if $L_1=L_2=\ldots=L_d$. Therefore, we conclude from \eqref{eq56} that at fixed
volume, the Casimir energy achieved its maximum when
$L_1=L_2=\ldots=L_d$.

\vspace{0.2cm}\noindent \textbf{A.2. High temperature expansion.}~
By putting $m=d$, $a_j =2\pi/L_j, 1\leq j\leq d$,
$a_{d+1}=2\pi/\beta$ in \eqref{eq25_10}, we obtain the high
temperature ($T=1/\beta\gg 1$) expansion of the free enrgy
\begin{align}\label{eq65}
&F_{P}(L_1, \ldots, L_d)=-\frac{\pi^{-\frac{d+1}{2}}}{\beta^{d+1}
}L_1\ldots L_d\Gamma\left(\frac{d+1}{2}\right)
\zeta_R(d+1)-\frac{1}{\beta}\log(2\pi\beta)\\&\hspace{2cm}-\frac{1}{2\beta}Z_{E,d}'(0;
L_1^{-1}, \ldots, L_d^{-1})\nonumber\\&-\frac{2L_1\ldots
L_d}{\beta^{\frac{d+2}{2}}}
\sum_{\mathbf{k}\in\Z^{d}\setminus\{\mathbf{0}\}}
\sum_{m=1}^{\infty}m^{\frac{d}{2}}\left(\sum_{j=1}^{d}\left[L_jk_j\right]^2\right)^{-\frac{d}{4}}
K_{\frac{d}{2}}\left(\frac{2\pi m}{\beta}
\sqrt{\sum_{j=1}^{d}\left[L_jk_j\right]^2}\right).\nonumber
\end{align}
The leading term
\begin{align}\label{eq25_6}-\frac{L_1\ldots
L_d}{\pi^{\frac{d+1}{2}}\beta^{d+1}
}\Gamma\left(\frac{d+1}{2}\right) \zeta_R(d+1)\end{align}is the
usual Stefan-Boltzmann term. In some of the existing literature
(e.g. \cite{OS}), the second leading term
$\frac{1}{\beta}\log(2\pi\beta)$ was overlooked. However, since this
term does not depend on the dimension of the space $L_1, \ldots,
L_d$, it does not contribute to the Casimir force. Nevertheless,
this term is essential for the validity of the thermodynamic
relation \eqref{eq5_23_7}. The last term in \eqref{eq65} is an
exponentially decay term. More precisely, it is bounded above by
$$\frac{2L_1\ldots
L_d}{\beta^{\frac{d+2}{2}}}\frac{c_{\frac{d}{2}}d\left(\left[\tfrac{d}{2}\right]\right)
!} {\min\left\{L_1^{\frac{d+1}{2}}, L_1^{\frac{2d+1}{2}}\right\}}
\frac{\left(1+e^{-\frac{2\pi L_1}{\beta\sqrt{d}}}\right)^{d-1}
}{\left(1-e^{-\frac{2\pi
L_1}{\beta\sqrt{d}}}\right)^{\left[\frac{d}{2}\right]+d+1}}e^{-\frac{2\pi
L_1}{\beta\sqrt{d}}}\sum_{l=0}^{\left[\frac{d}{2}\right]}\beta^{l+\frac{1}{2}}.$$
Ambj{\o}rn and Wolfram obtained a similar high temperature expansion
in \cite{AW} (see (7.10)). They   considered the non-closed cavity
case and let $p=d$ in the formula valid for $p<d$, and then removed
the divergent term by subtracting the free bose gas result. They did
not justify their result mathematically. Here we have proved this
formula rigorously.

We would also like to mention that the general structure of the high
temperature expansion of free energy of gases inside cavities in
curved space--time has been calculated (see e.g. \cite{DK, DS,
Kie}). Our result here can be considered as special case of their
result.

 \vspace{0.5cm}\noindent \textbf{B) Dirichlet and Neumann Boundary
Condition}

\vspace{0.2cm}\noindent \textbf{B.1. Dirichlet Boundary Condition.}~
 The eigenmodes of $\phi$ satisfying the
Dirichlet boundary condition
$\left.\phi(\mathbf{x})\right|_{\pa\Omega}=0$ are
\begin{align*}
\phi_{\mathbf{k}}(\mathbf{x})= \prod_{j=1}^d \sin\left(\frac{\pi
k_j}{L_j}x_j\right), \hspace{1cm} \mathbf{k}\in \mathbb{N}^d.
\end{align*}The corresponding zeta function $\zeta(s)$ is
\begin{align*}
\zeta_{D, d}(s; L_1,\ldots,L_d) =&\sum_{(m,\mathbf{k})\in \Z\times
\mathbb{N}^d}\left(\left(\frac{2\pi
m}{\beta}\right)^2+\left(\frac{\pi k_1}{L_1}\right)^2+\ldots
+\left(\frac{\pi k_d }{L_d}\right)^2\right)^{-s}.
\end{align*}There is no zero mode of $\phi$ in this case.

\vspace{0.2cm}\noindent \textbf{B.2. Neumann Boundary Condition.} ~
For the Neumann boundary condition
$\left.\pa_{\mathbf{n}}\phi(\mathbf{x})\right|_{\pa\Omega}=0$, where
$\mathbf{n}$ denotes the unit vector normal to the surface
$\pa\Omega$, the eigenmodes of $\phi$ are
\begin{align*}
\phi_{\mathbf{k}}(\mathbf{x})=\prod_{j=1}^d \cos\left(\frac{\pi
k_j}{L_j}x_j\right), \hspace{1cm} \mathbf{k}\in
\left(\mathbb{N}\cup\{0\}\right)^d.
\end{align*}The corresponding zeta function $\zeta(s)$ is
\begin{align*}
\zeta_{N, d}(s; L_1,\ldots,L_d) =&\sum_{(m,\mathbf{k})\in \Z\times
\left(\mathbb{N}\cup\{0\}\right)^d}\!'\left(\left(\frac{2\pi
m}{\beta}\right)^2+\left(\frac{\pi k_1}{L_1}\right)^2+\ldots
+\left(\frac{\pi k_d }{L_d}\right)^2\right)^{-s}.
\end{align*}There is $\mathcal{N}=1$ zero mode of $\phi$ in this case corresponding to $\mathbf{k}=\mathbf{0}$.

\vspace{0.2cm}

\noindent Since
\begin{align*}
\sum_{\mathbf{k}\in\mathbb{N}^d} g(k_1, \ldots,
k_d)=&2^{-d}\sum_{\mathbf{k}\in\mathbb{Z}^d}\left(1-\delta_{k_1,
0}\right)\ldots\left(1-\delta_{k_d,0}\right)g(k_1, \ldots,k_d),\\
\sum_{\mathbf{k}\in\left(\mathbb{N}\cup\{0\}\right)^d}  g(k_1,
\ldots,
k_d)=&2^{-d}\sum_{\mathbf{k}\in\mathbb{Z}^d}\left(1+\delta_{k_1,
0}\right)\ldots\left(1+\delta_{k_d,0}\right)g(k_1, \ldots,k_d)
\end{align*}for any function $g$ satisfying $g(k_1,\ldots, -k_i,
\ldots, k_d)=g(k_1, \ldots, k_i, \ldots, k_d)$, $1\leq i\leq d$, we
have
\begin{align*}
&\zeta_{D/N, d}(s; L_1,\ldots,L_d)=2^{-d}\Biggl(2(\mp
1)^d\left(\frac{2\pi}{\beta}\right)^{-2s}\zeta_R(2s)\\&
\hspace{2cm}+\sum_{j=1}^{d}(\mp 1)^{d-j} \sum_{1\leq
m_1<\ldots<m_j\leq d}Z_{E,j+1}\left(s;
\frac{2\pi}{\beta},\frac{\pi}{L_{m_1}},
\ldots,\frac{\pi}{L_{2m_j}}\right)\Biggr).
\end{align*}From this, it is easy to check that $\zeta_{D, d}(0;
L_1,\ldots,L_d)=0$ and  $\zeta_{N, d}(0; L_1,\ldots,L_d)=-1$.
Therefore, by \eqref{eq12} the free energy is given by
\begin{align}\label{eq61}
F_{D/N}(L_1, \ldots, L_d)=&-\frac{1}{2^{d+1}\beta}\sum_{j=1}^{d}(\mp
1)^{d-j} \sum_{1\leq m_1<\ldots<m_j\leq d}Z_{E,j+1}'\left(0;
\frac{2\pi}{\beta},\frac{\pi}{L_{m_1}},
\ldots,\frac{\pi}{L_{m_j}}\right)\\
&+\frac{(\mp
1)^d}{2^d\beta}\log\beta-\theta_{D/N}\frac{1}{\beta}\log\beta\nonumber,
\end{align}where $\theta_D=0$ and $\theta_N=1$.
Compare to the free energy of the periodic case \eqref{eq30_3}, we
have
\begin{align}\label{eq3_1}
F_{D/N}(L_1, \ldots, L_d)=&2^{-d}\sum_{j=1}^{d}(\mp 1)^{d-j}
\sum_{1\leq m_1<\ldots<m_j\leq d}F_p(2L_{m_1}, \ldots, 2L_{m_j}).
\end{align}
Using  this formula and \eqref{eq17_3}, we find that under the
simultaneous space--time scaling $\beta\mapsto \lambda \beta$,
$L_i\mapsto \lambda L_i, 1\leq i\leq d$, the free energy for the
Dirichlet and Neumann conditions $F_{D/N}(L_1, \ldots, L_d)$ behave
in the same way as the free energy for the periodic condition
$F_P(L_1, \ldots, L_d)$ \eqref{eq17_3}, and thus the thermodynamic
relation \eqref{eq5_23_7} also holds in these cases.

\begin{figure}\centering \epsfxsize=.45\linewidth
\epsffile{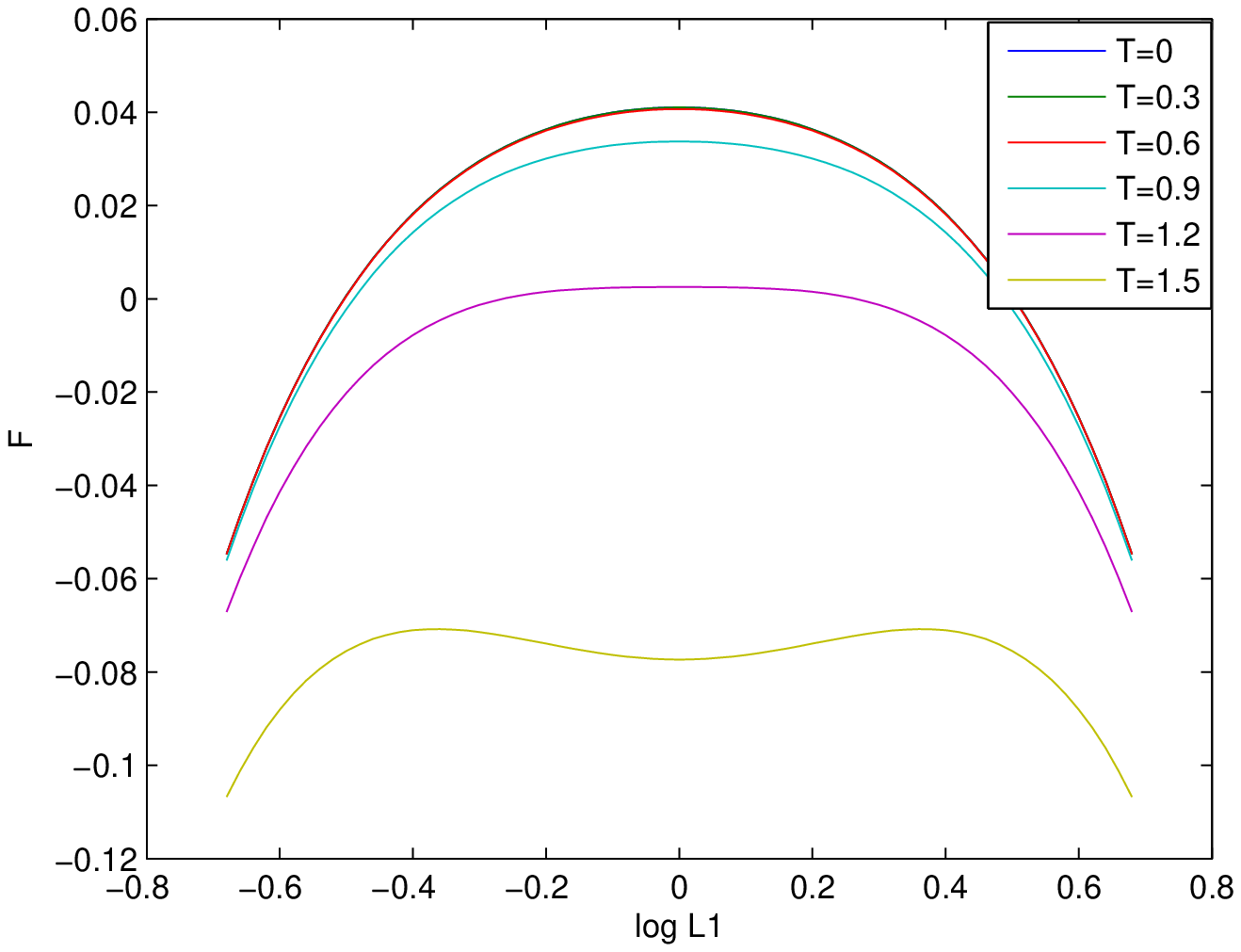}\epsfxsize=.45\linewidth
\epsffile{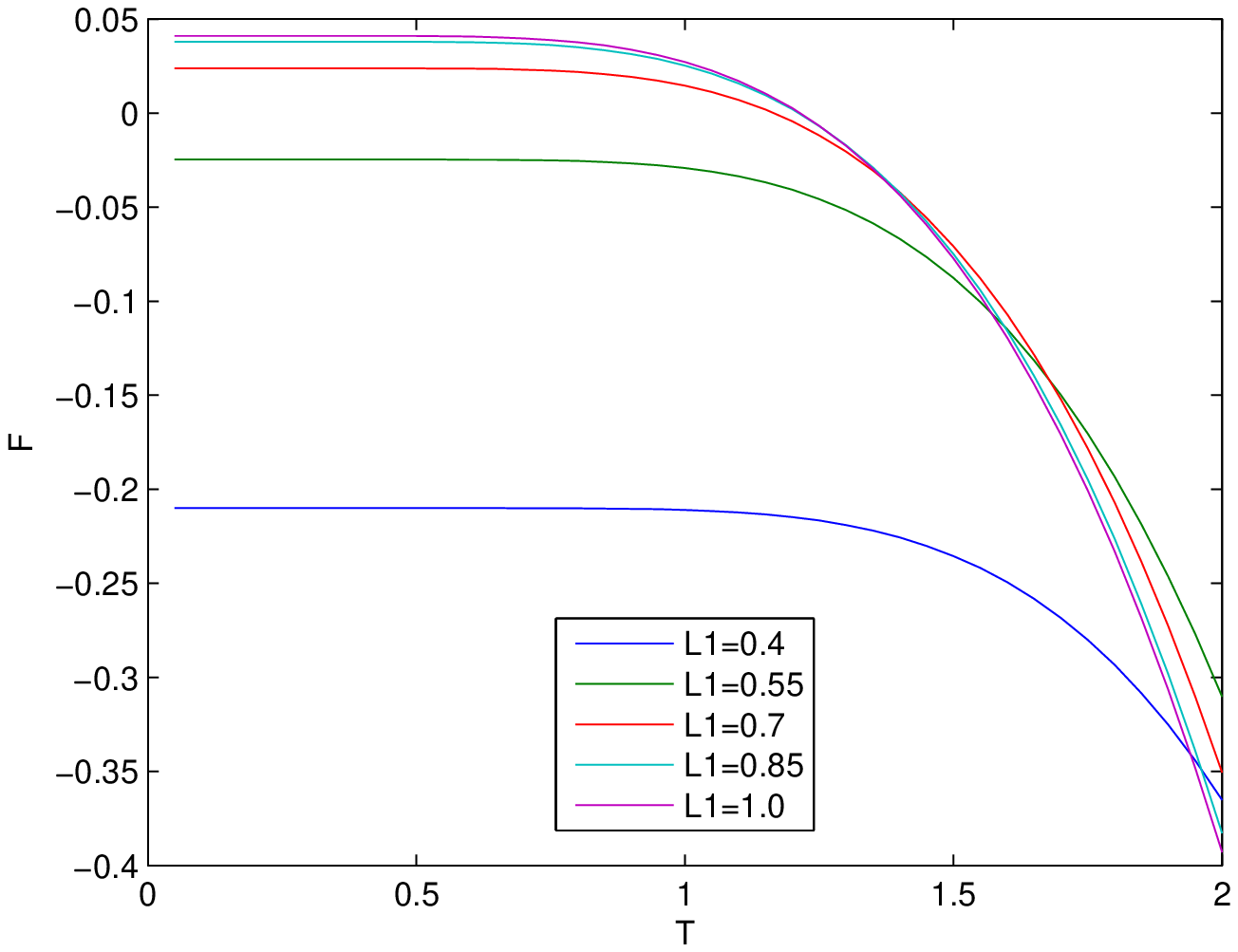}\caption{The graph on the left shows the
free energy $F_D(L_1, L_2)$ as a function of $L_1$ when
$V=L_1L_2=1$, at $T=0,0.3,0.6,0.9,1.2,1.5$.
 The graph on the right shows the free energy $F_D(L_1, L_2)$ as a function of $T$ when $L_1=0.4,0.55,
 0.7,0.85,1.0$ and $V=L_1L_2=1$.}\end{figure}

\begin{figure}\centering \epsfxsize=.40\linewidth
\epsffile{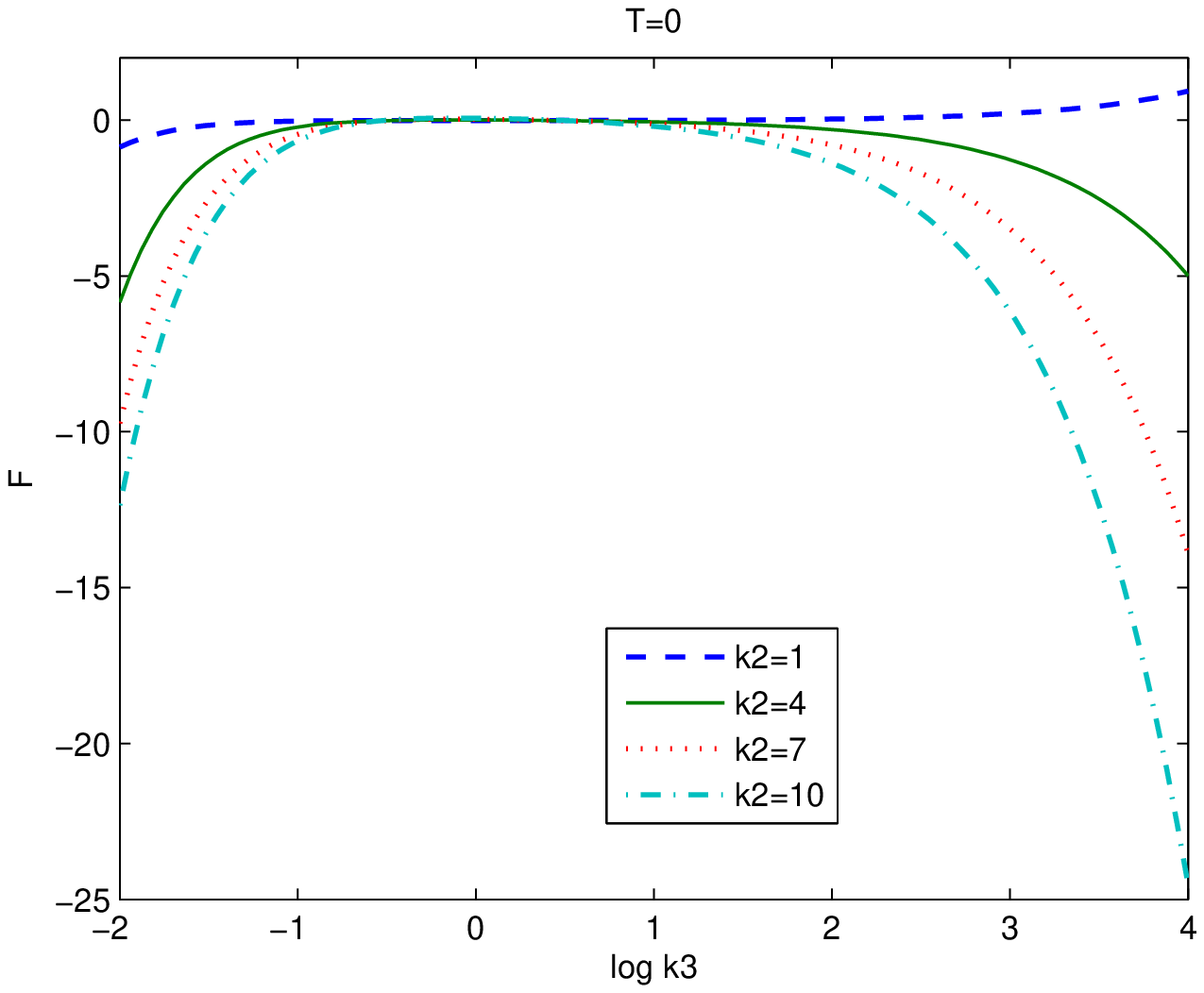}\epsfxsize=.40\linewidth
\epsffile{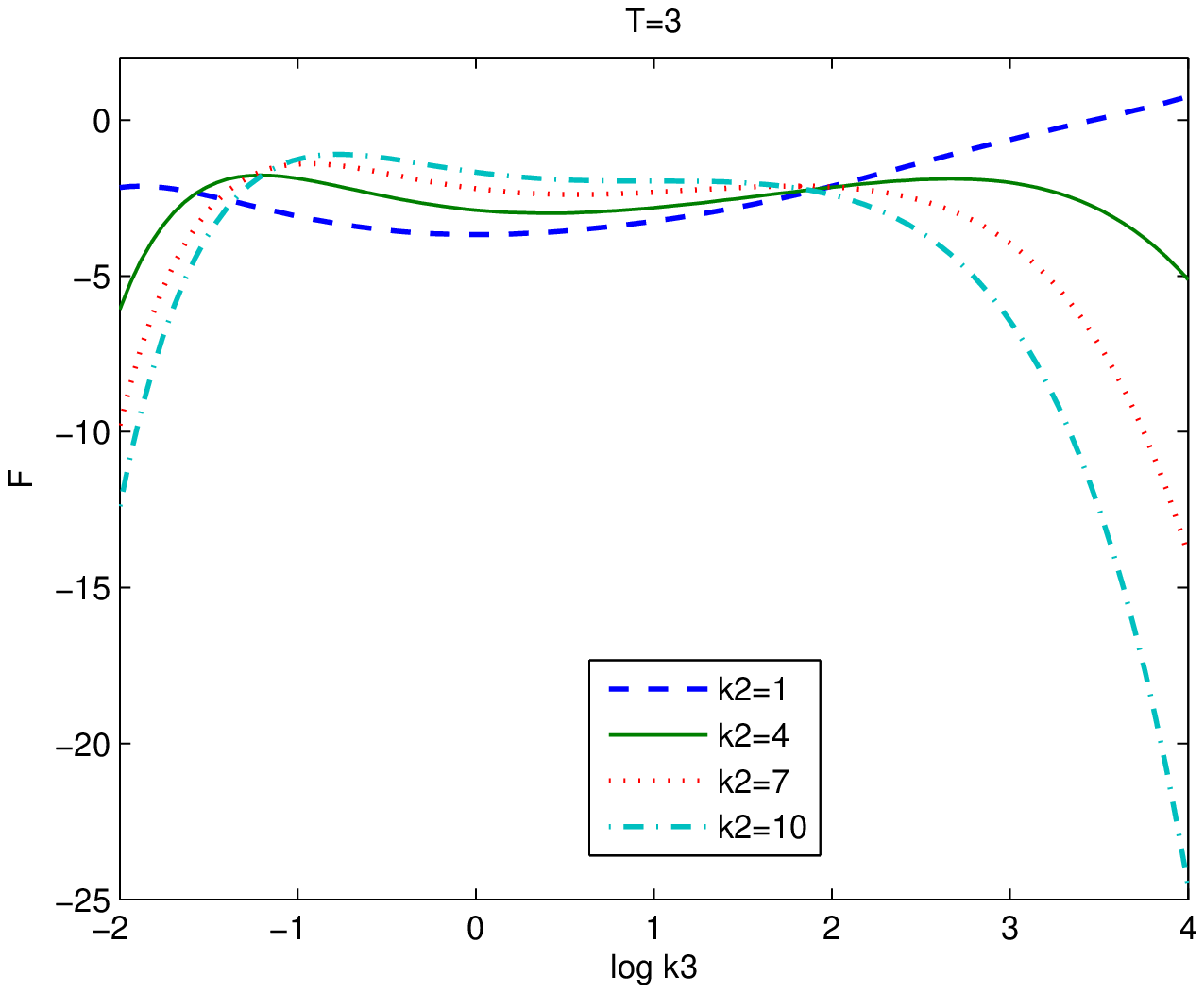}\\\epsfxsize=.40\linewidth
\epsffile{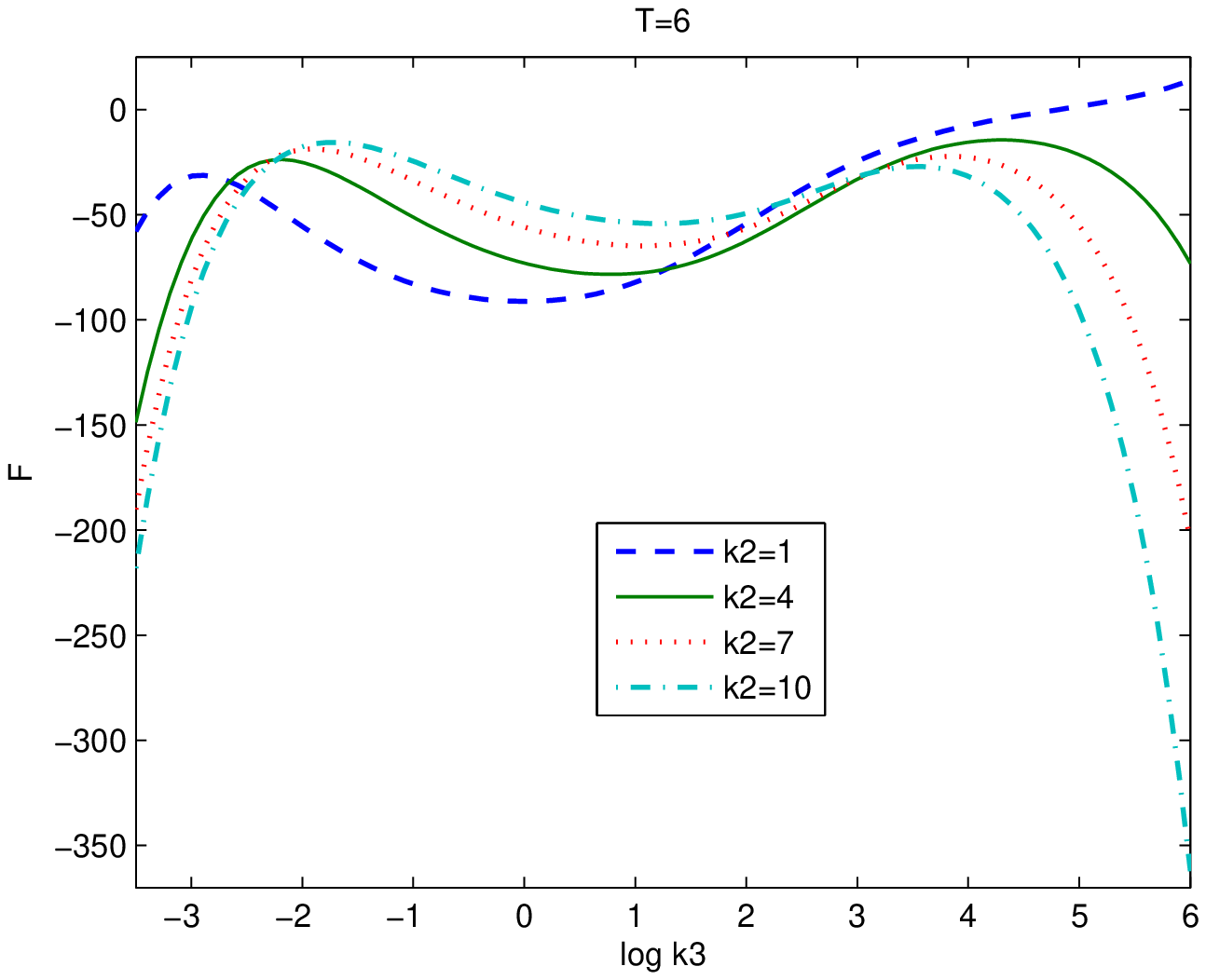}\epsfxsize=.40\linewidth
\epsffile{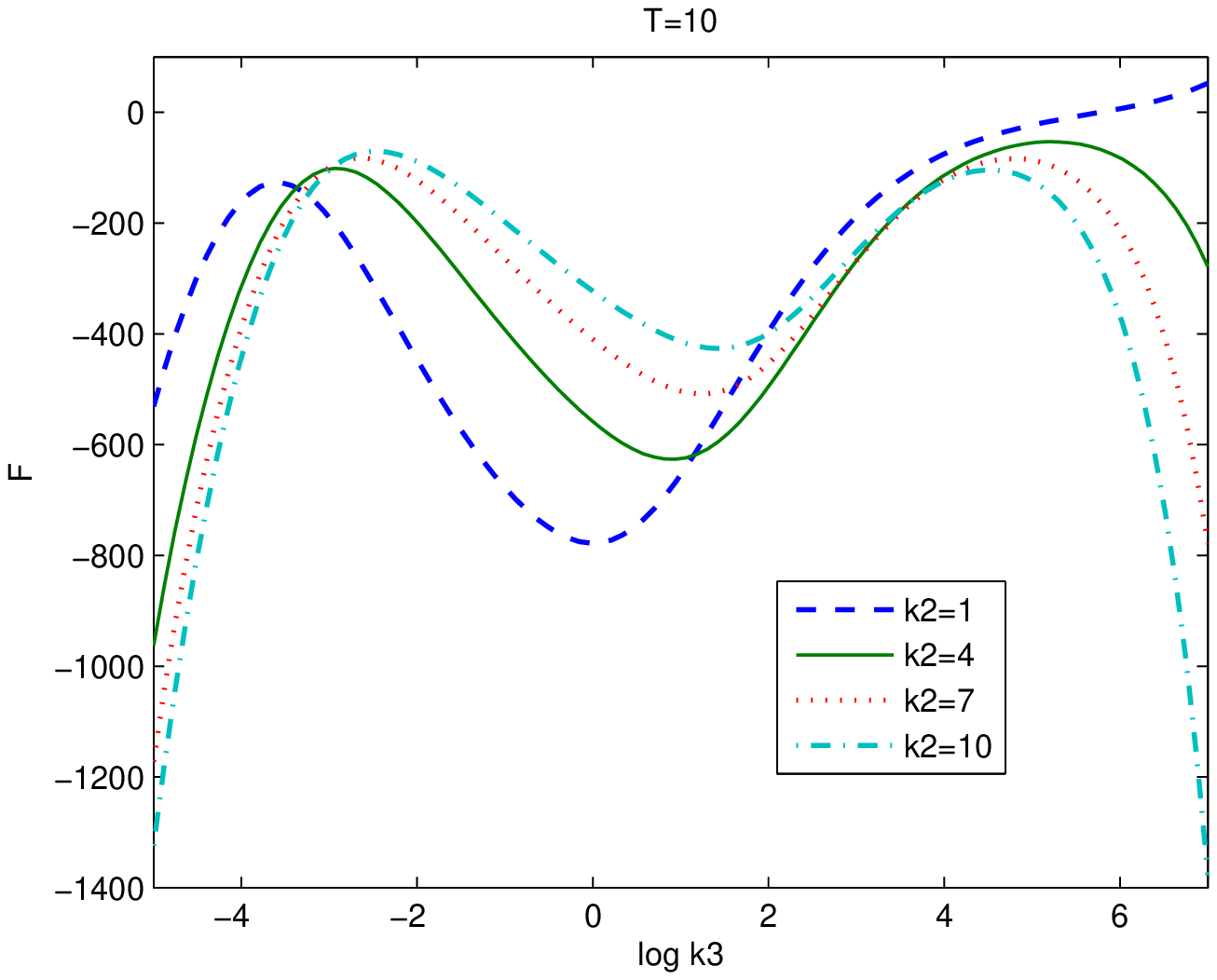}\caption{The free energy $F_D(L_1, L_2, L_3)$
as a function of $k_3=L_3/L_1$ when  $k_2=L_2/L_1=1, 4, 7, 10$
 and $V=L_1L_2L_3=1$, at $T=0,3,6,10$ respectively.
}\end{figure}

\begin{figure}\centering \epsfxsize=.40\linewidth
\epsffile{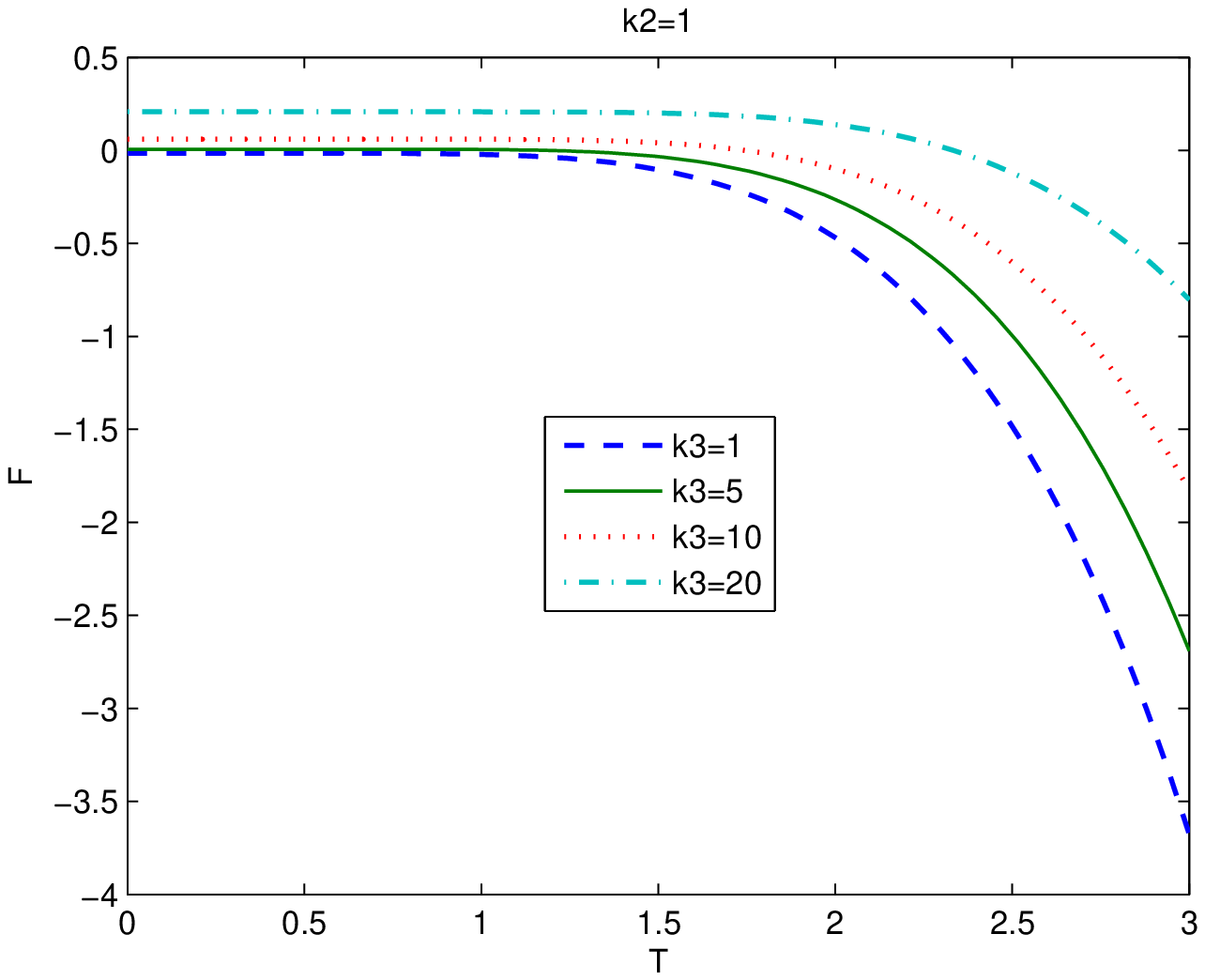}\epsfxsize=.40\linewidth
\epsffile{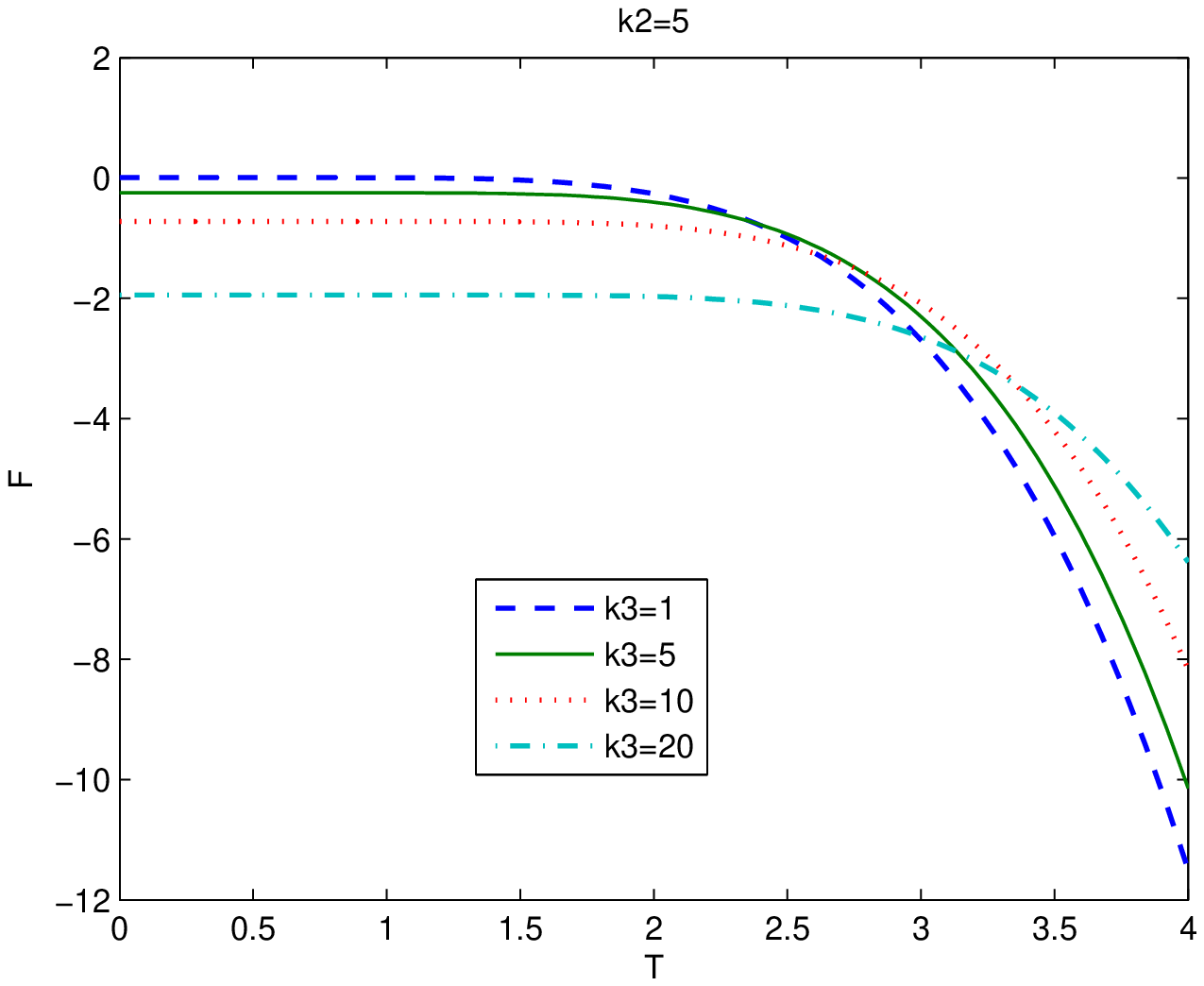}\caption{The free energy $F_D(L_1, L_2, L_3)$
as a function of $T$ when $V=L_1L_2L_3=1$ and $k_2=L_2/L_1=1, 5$ at
 $k_3=L_3/L_1=1, 5, 10,20$ respectively. }\end{figure}

\begin{figure}\centering \epsfxsize=.40\linewidth
\epsffile{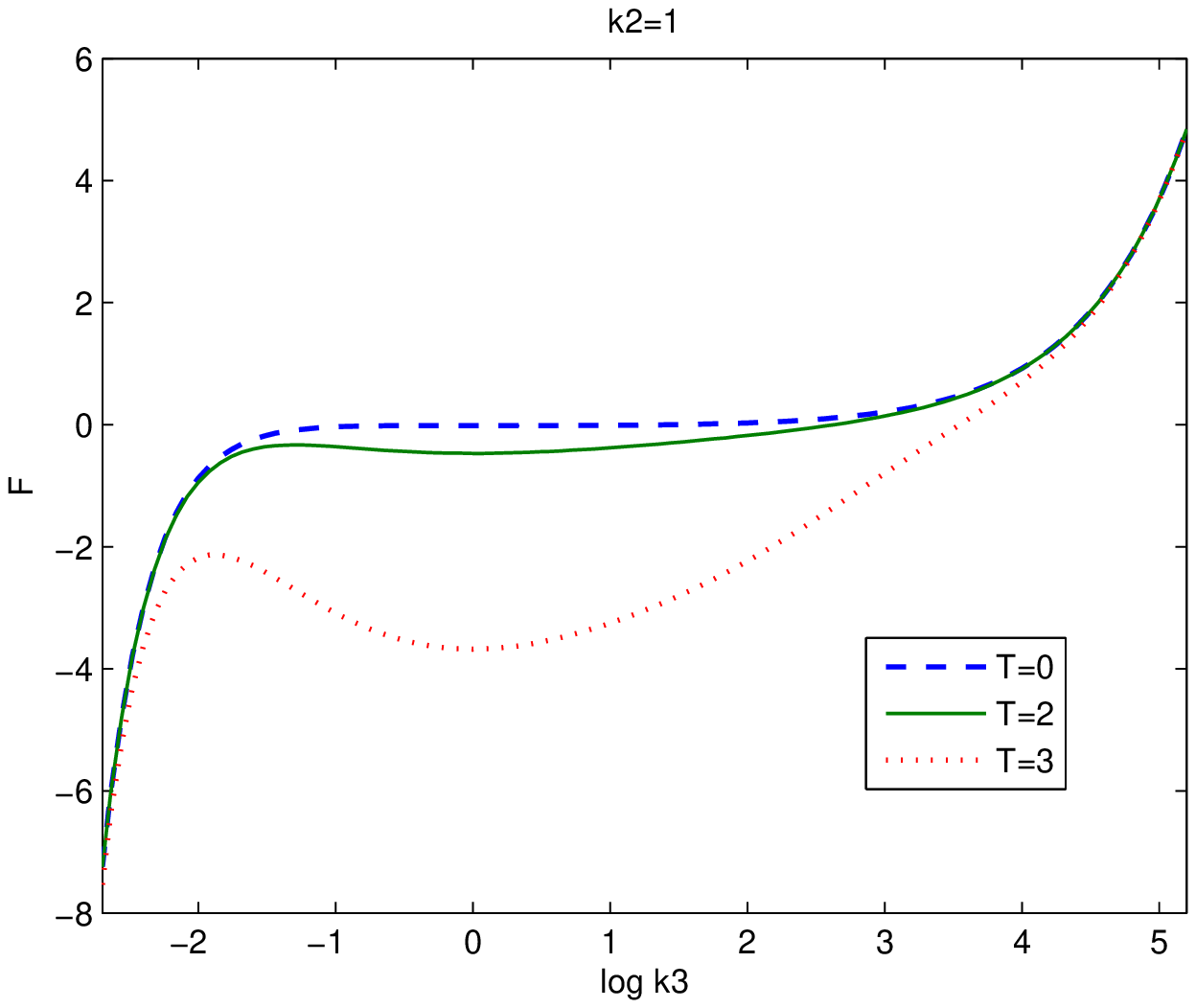}\epsfxsize=.40\linewidth
\epsffile{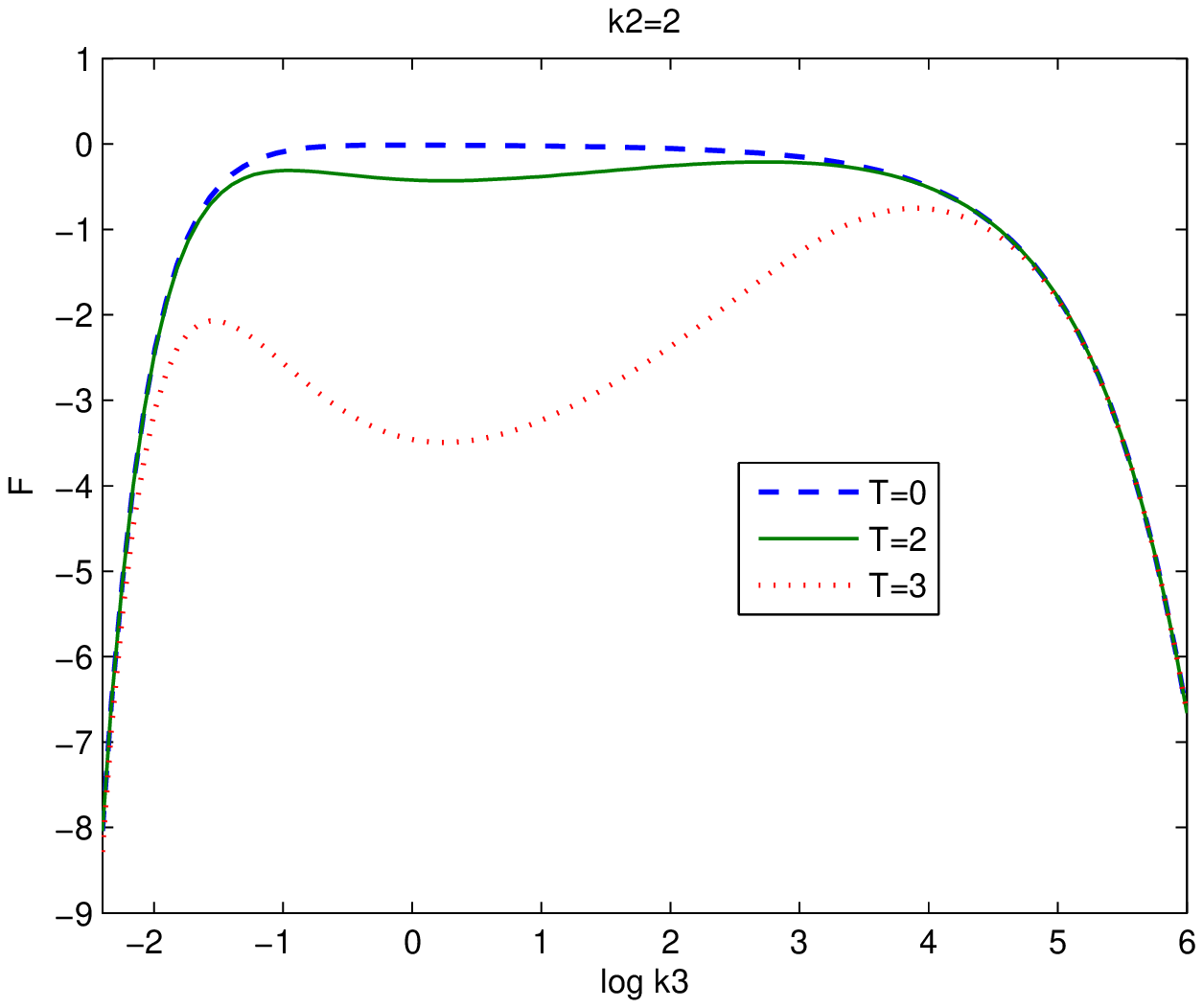}\\\epsfxsize=.40\linewidth
\epsffile{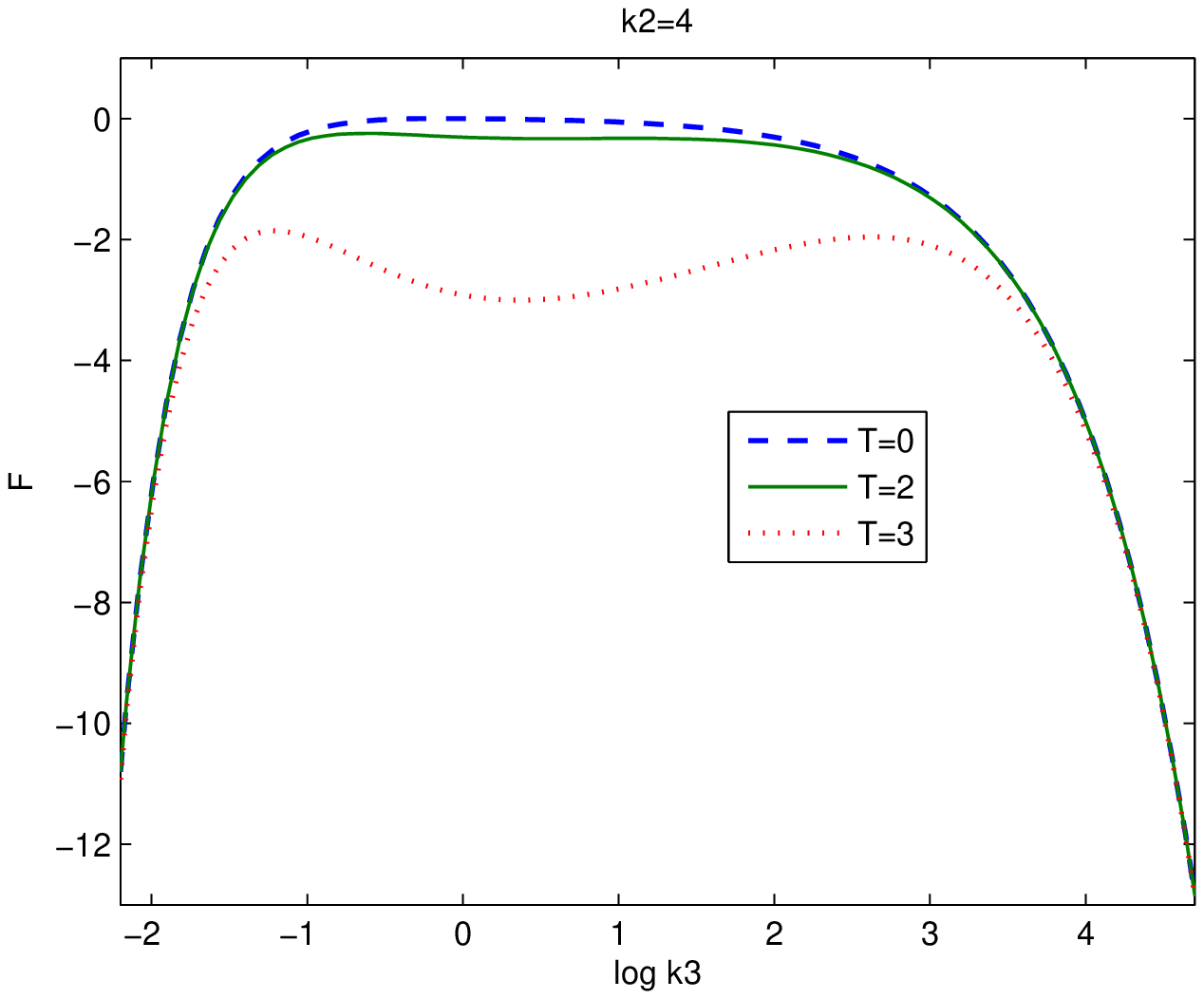}\epsfxsize=.40\linewidth
\epsffile{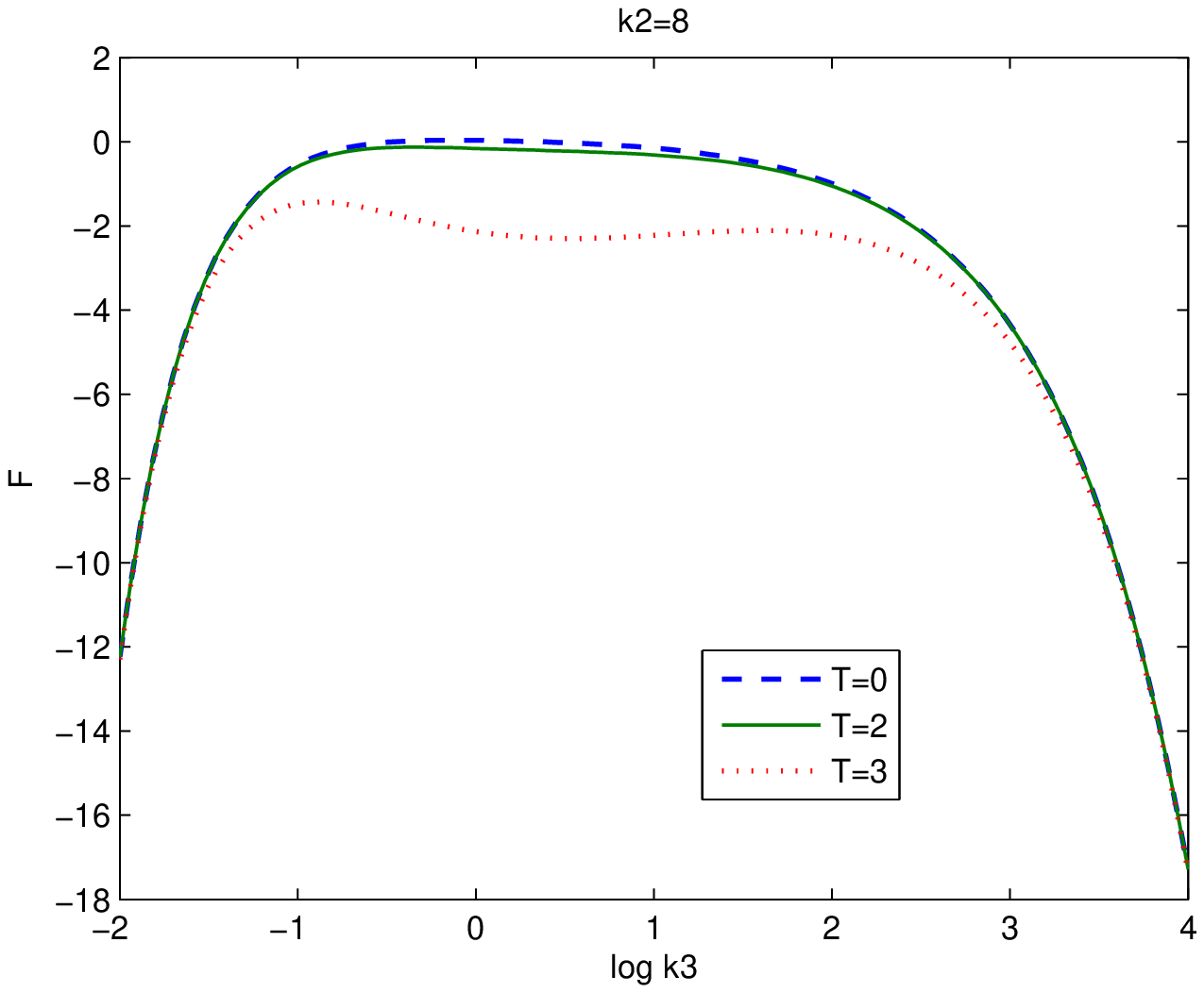}\caption{The free energy $F_D(L_1, L_2, L_3)$
as a function of $k_3=L_3/L_1$ when  $T=0, 2, 3$  and
$V=L_1L_2L_3=1$ at $k_2=L_2/L_1=1,2,4,8$.}\end{figure}

The low and high temperature expansions of the Casimir free energy
$F_{D/N}$
 can be obtained directly from \eqref{eq3_1} and the corresponding
 expansion for $F_P$.
 As in the
periodic case, the zero temperature free energy for the Neumann case
is always negative. However, the sign of the zero temperature free
energy of the Dirichlet case depends on the parameters $L_1, \ldots,
L_d$. There have been a lots of discussions about this in the
literature, see e.g. \cite{Ch1, AW,  CNS, Mac}. By Remark \ref{re1},
we know that fixing $L_1, \ldots, L_d$, if $F^0_D$ is positive, then
there exists a unique $T=T(L_1, \ldots, L_d)$ such that $F_D(L_1,
\ldots, L_D)$ change from negative to positive.  The scaling
property  of free energy \eqref{eq17_3} shows that $T(\lambda L_1,
\ldots, \lambda L_d) = \lambda^{-1}T(L_1, \ldots, L_d)$. We study
this transition temperature graphically for $d=2$ and $d=3$ (see
Figure 5, 6).

\vspace{0.5cm} \noindent\textbf{Table 1}
\\
\begin{tabular}{c|c}
\hline
$k_2=L_2/L_1$ & The range of $k_3=L_3/L_1$ where $F_{D}^0(L_1, L_2, L_3)\geq 0$\\
\hline
0.75 &   $k_3\geq  4.4972$\\
\hline 1.0&  $k_3\geq 4.2471    $  \\
\hline
 1.25&  $k_3\geq  5.2999$\\ \hline
1.5&   $k_3\geq 8.8571$ \\
\hline
\end{tabular}
\\
\\

\begin{figure}\centering \epsfxsize=.45\linewidth
\epsffile{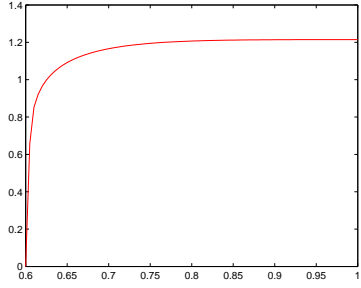}\caption{The transition temperature
$T(L_1, L_2)$ for $F_D(L_1, L_2)$ as a function of $L_1$ when
$V=L_1L_2=1$. $F_D^0(L_1, L_2)$ is positive when $0.6045^2\leq
L_1/L_2 \leq 1/0.6045^2$.} \end{figure}

\begin{figure}\centering
\epsfxsize=.45\linewidth \epsffile{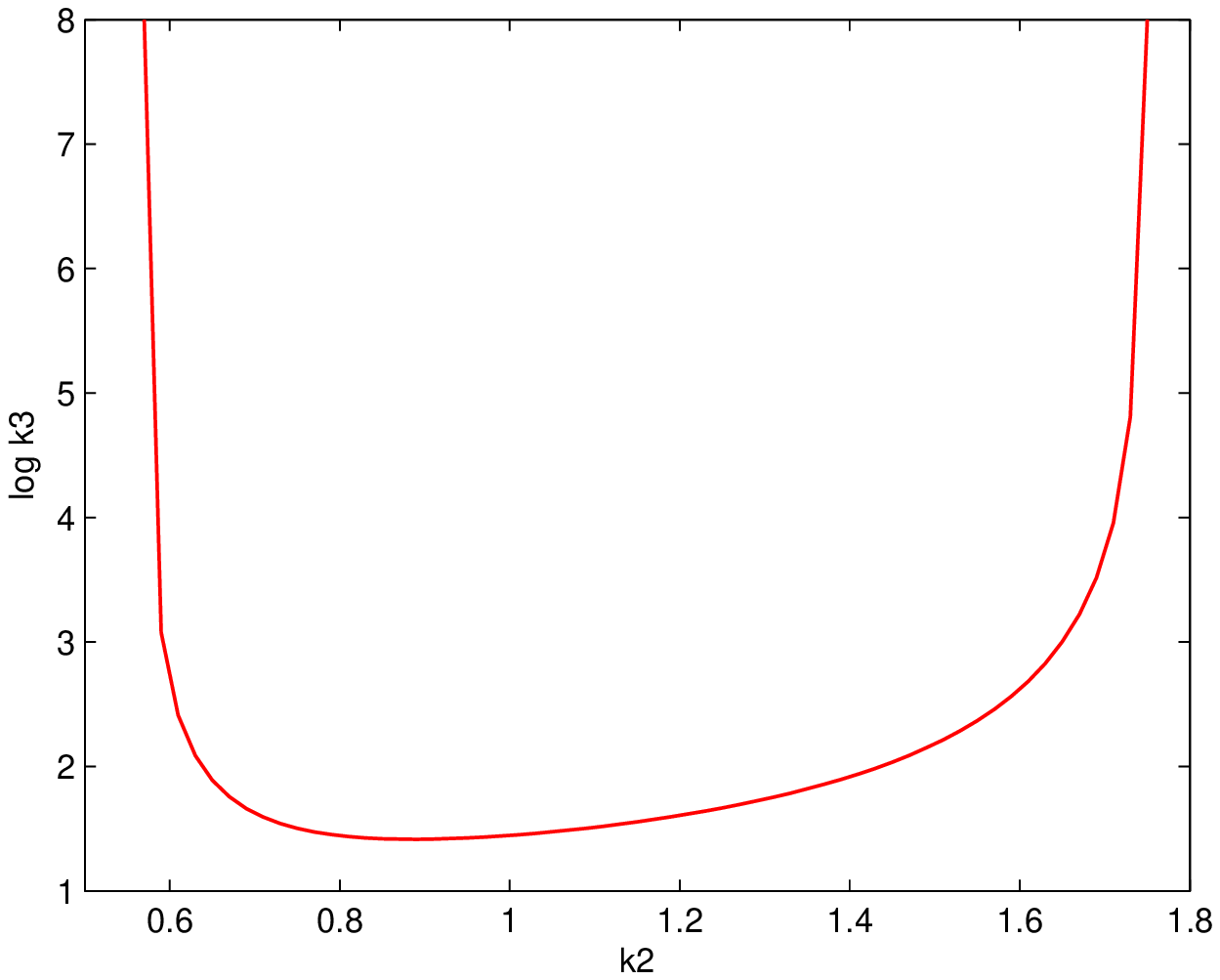}
 \epsfxsize=.45\linewidth
\epsffile{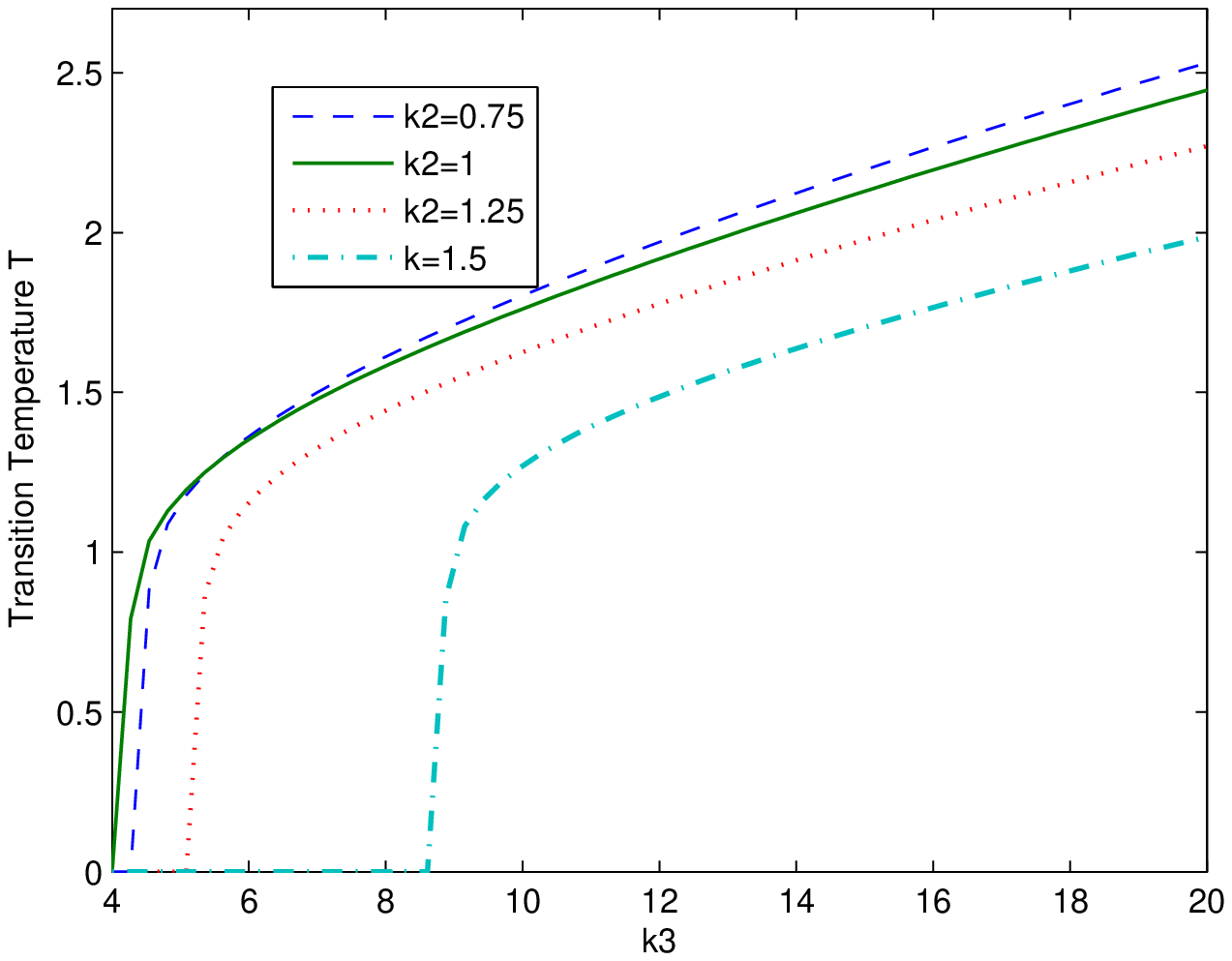}\caption{\textbf{Left:} When
$0.5733\leq k_2=L_2/L_1\leq 1.7444$, there is a unique
$\hat{k}_3=\hat{k}_3(k_2)$ such that $F_D(L_1,L_2, L_3)\geq 0$ for
all $L_3/L_1\geq \hat{k}_3$.  The graph shows $\log \hat{k}_3$ as a
function of $k_2$. \textbf{Right:} The transition temperature
$T(L_1, L_2,L_3)$ for $F_D(L_1, L_2,L_3)$ as a function of
$k_3=L_3/L_1$ when $V=1$ and $k_2=L_2/L_1=0.75,1, 1.25, 1.5$. }
\end{figure}

In the high temperature regime, the leading term is
$$-\frac{L_1\ldots L_d}{\pi^{\frac{d+1}{2}}\beta^{d+1}
}\Gamma\left(\frac{d+1}{2}\right) \zeta_R(d+1).$$ It comes from the
$j=d$ term in \eqref{eq3_1} and it is again the Stefan-Boltzman term
 as in the periodic case \eqref{eq25_6}.  The term proportional
  to $\frac{1}{\beta}\log\beta$ is also present
but in the Dirichlet case, its
 sign depends on $d$. Unlike the
periodic boundary case \eqref{eq65}, now we have terms proportional
to $1/\beta^j$ for all $1\leq j\leq d+1$.

\section{Massless vector field (electromagnetic field)}

As discussed in \cite{AW}, for massless vector (spin 1) field (or
electromagnetic field) inside a $d$-dimensional space $\Omega$, the
field strength is represented by a totally anti-symmetric rank-2
tensor $F^{\mu\nu}$ satisfying the equations
\begin{align*}
  \pa_{\mu}\tilde{F}^{\mu\nu_1\ldots\nu_{d-2}}=&0,\\
 \pa_{\mu}F^{\mu\nu}=&j^{\nu},
\end{align*}where $\tilde{F}^{\mu_1\ldots
\mu_{d-1}}=\varepsilon^{\mu_1,\ldots,\mu_{d-1},\nu,\lambda}F_{\nu\lambda}$
is the dual tensor of $F^{\mu\nu}$ and $j^{\mu}$ is the current. In
the vacuum state, $j^{\mu}=0$.

\vspace{0.2cm} \noindent \textbf{C.1. Perfectly Conducting Walls.}~~
 In the case that $\Omega=[0,
L_1]\times\ldots\times[0, L_d]$ is a rectangular cavity with walls
of infinite conductivity, the field satisfies the boundary condition
\begin{align*}
\left.n_{\mu}\tilde{F}^{\mu\nu_1\ldots\nu_{d-1}}\right|_{\pa\Omega}=0,
\end{align*}where $n_{\mu}$ is the unit  vector normal to the walls $\pa\Omega$ and $n_0=0$.
Introducing the potentials $A^{\mu}$ so that
\begin{align*}
F^{\mu\nu}=\pa^{\mu}A^{\nu}-\pa^{\nu}A^{\mu},\hspace{1cm}\pa^0=\pa_0,
\pa^i=-\pa_i, \;\;1\leq i\leq d
\end{align*}and working in the radiation  gauge with gauge condition
\begin{align}\label{eq27_1}
A^0=&0, \\\pa_iA^i=&0,\nonumber
\end{align}we find that the modes of the potentials  are given by
\begin{align*}
A_{\mathbf{k}}^{i} =& \alpha_i\cos\left(\frac{\pi
k_i}{L_i}x_i\right)\prod_{\substack{j=1\\j\neq
i}}^d\sin\left(\frac{\pi k_j}{L_j}x_j\right)e^{-i\omega_{\mathbf{k}}
t},\hspace{1cm} 1\leq i\leq d,\end{align*}
$$\text{where}\hspace{1cm}\mathbf{k}\in (\mathbb{N}\cup\{0\})^d,\hspace{1cm}\omega_{\mathbf{k}}=\sqrt{\sum_{j=1}^d\left(\frac{\pi
k_j}{L_j}\right)^2}.$$ The gauge condition \eqref{eq27_1} implies
\begin{align}\label{eq27_2}\sum_{i=1}^d \frac{\alpha_i k_i}{L_i}=0.\end{align} It is easy to see
that if two of the $k_i$'s are zero,
$\mathbf{A}_{\mathbf{k}}=(A_{\mathbf{k}}^0,\ldots,A_{\mathbf{k}}^d)$
is identically 0. On the other hand, if only a single $k_i$ is zero,
then for $j\neq i$,    $A_{\mathbf{k}}^j=0$ and \eqref{eq27_2} is
trivially satisfied. When all $k_i$'s are nonzero, \eqref{eq27_2}
implies that there is a $(d-1)$ degree of freedom for the vector
$\vec{\alpha}=(\alpha_1, \ldots,\alpha_d)$ for any fixed
$\mathbf{k}\in \mathbb{N}^d$.  Therefore the  zeta function for
electromagnetic field confined in rectangular cavities with
perfectly conducting walls is related to the zeta function for
massless scalar field under Dirichlet boundary condition by
\begin{align*}
\zeta_{A_C, d}(s; L_1, \ldots, L_d)=&(d-1)\zeta_{D,d}(s; L_1,
\ldots, L_d)\\&+\sum_{j=1}^{d} \zeta_{D,d-1}(s; L_1, \ldots,
L_{j-1}, L_{j+1}, \ldots, L_d).
\end{align*}There is no $\omega_{\mathbf{k}}=0$ mode and $\zeta_{A_C, d}(0; L_1, \ldots,
L_d)=0$. The corresponding free energy is
\begin{align}\label{eq27_3}
F_{A_C}(L_1, \ldots, L_d)=(d-1)F_D(L_1,\ldots,L_d)+\sum_{j=1}^d
F_{D}(L_1,\ldots, L_{j-1}, L_{j+1}, \ldots, L_d)\\
=2^{-d} \sum_{j=1}^{d}(- 1)^{d-j}(2j-d-1) \sum_{1\leq
m_1<\ldots<m_j\leq d}F_p(2L_{m_1}, \ldots, 2L_{m_j})\nonumber.
\end{align}

\vspace{0.2cm} \noindent \textbf{C.2. Infinitely Permeable Walls.}~
In the case that $\Omega=[0, L_1]\times\ldots\times[0, L_d]$ is a
rectangular cavity where the walls are infinitely permeable, the
field satisfies the boundary condition
\begin{align*}
\left.n_{\mu}F^{\mu\nu}\right|_{\pa\Omega}=0.
\end{align*}Working in the  radiation gauge \eqref{eq27_1},
 the modes of the potentials  are given by
\begin{align*}
A_{\mathbf{k}}^{i} =& \gamma_i\sin\left(\frac{\pi
k_i}{L_i}x_i\right)\prod_{\substack{j=1\\j\neq
i}}^d\cos\left(\frac{\pi k_j}{L_j}x_j\right)e^{-i\omega_{\mathbf{k}}
t},\hspace{1cm} 1\leq i\leq d,\end{align*}
$$\text{where}\hspace{2cm}\mathbf{k}\in (\mathbb{N}\cup\{0\})^d,\hspace{1cm}\omega_{\mathbf{k}}=\sqrt{\sum_{j=1}^d\left(\frac{\pi
k_j}{L_j}\right)^2}.$$The gauge condition \eqref{eq27_1} implies
\begin{align}\label{eq17_5}\sum_{i=1}^d \frac{\gamma_i k_i}{L_i}=0.\end{align} If all the $k_i$'s are zero,
$\mathbf{A}_{\mathbf{k}}=(A_{\mathbf{k}}^0,\ldots,A_{\mathbf{k}}^d)$
is identically 0. On the other hand, for $1\leq j\leq d$, fixing
$1\leq r_1<\ldots< r_{d-j}\leq d$, let $1\leq m_1<\ldots <m_j\leq d$
be such that $\{m_1,\ldots,m_j, r_1, \ldots,r_{d-j}\}=\{1,2,\ldots,
d\}$. If $k_{r_1}=\ldots k_{r_{d-j}}=0$  and $k_{m_1}\neq
0,\ldots,k_{m_j}\neq 0$, then $A_{\mathbf{k}}^{r_l}=0$ for $1\leq
l\leq d-j$ and \eqref{eq17_5} reduces to $\sum_{l=1}^{
j}\frac{\gamma_{m_l} k_{m_l}}{L_{m_l}}=0$. This implies that there
is a $(j-1)$ degrees of freedom for the vector $(\gamma_{m_1},
\ldots,\gamma_{m_{j}})$ for any fixed $(k_{m_1},\ldots, k_{m_j}) \in
\mathbb{N}^j$. Therefore the zeta function for electromagnetic field
confined in a closed rectangular cavity with infinitely permeable
walls is related to the zeta function for massless scalar field
under Dirichlet boundary condition by
\begin{align*}
\zeta_{A_B, d}(s; L_1, \ldots, L_d)=&\sum_{j=2}^{d}(j-1)\sum_{1\leq
m_1<\ldots <m_j\leq d}\zeta_{D,j}\left(s; L_{m_1},\ldots,
L_{m_j}\right).
\end{align*}There is no $\omega_{\mathbf{k}}=0$ mode and $\zeta_{A_B, d}(0; L_1, \ldots,
L_d)=0$. The corresponding free energy is (see the detail
computation in the appendix):
\begin{align}\label{eq27_3_1}
F_{A_B}(L_1, \ldots, L_d)=&\sum_{j=2}^{d}(j-1)\sum_{1\leq m_1<\ldots
<m_j\leq d}F_{D}\left(s; L_{m_1},\ldots,
L_{m_j}\right).\\
=&2^{-d} \sum_{j=1}^{d}(2j-d-1) \sum_{1\leq m_1<\ldots<m_j\leq
d}F_p(2L_{m_1}, \ldots, 2L_{m_j})\nonumber.
\end{align}Notice that when $d=2$, $F_{A_C}(L_1,L_2)=F_N(L_1,L_2)$, $F_{A_B}(L_1,L_2)=F_D(L_1,L_2)$ and when $d=3$, $F_{A_C}(L_1, L_2,
L_3)=F_{A_B}(L_1, L_2, L_3)$.

\begin{figure}\centering \epsfxsize=.45\linewidth
\epsffile{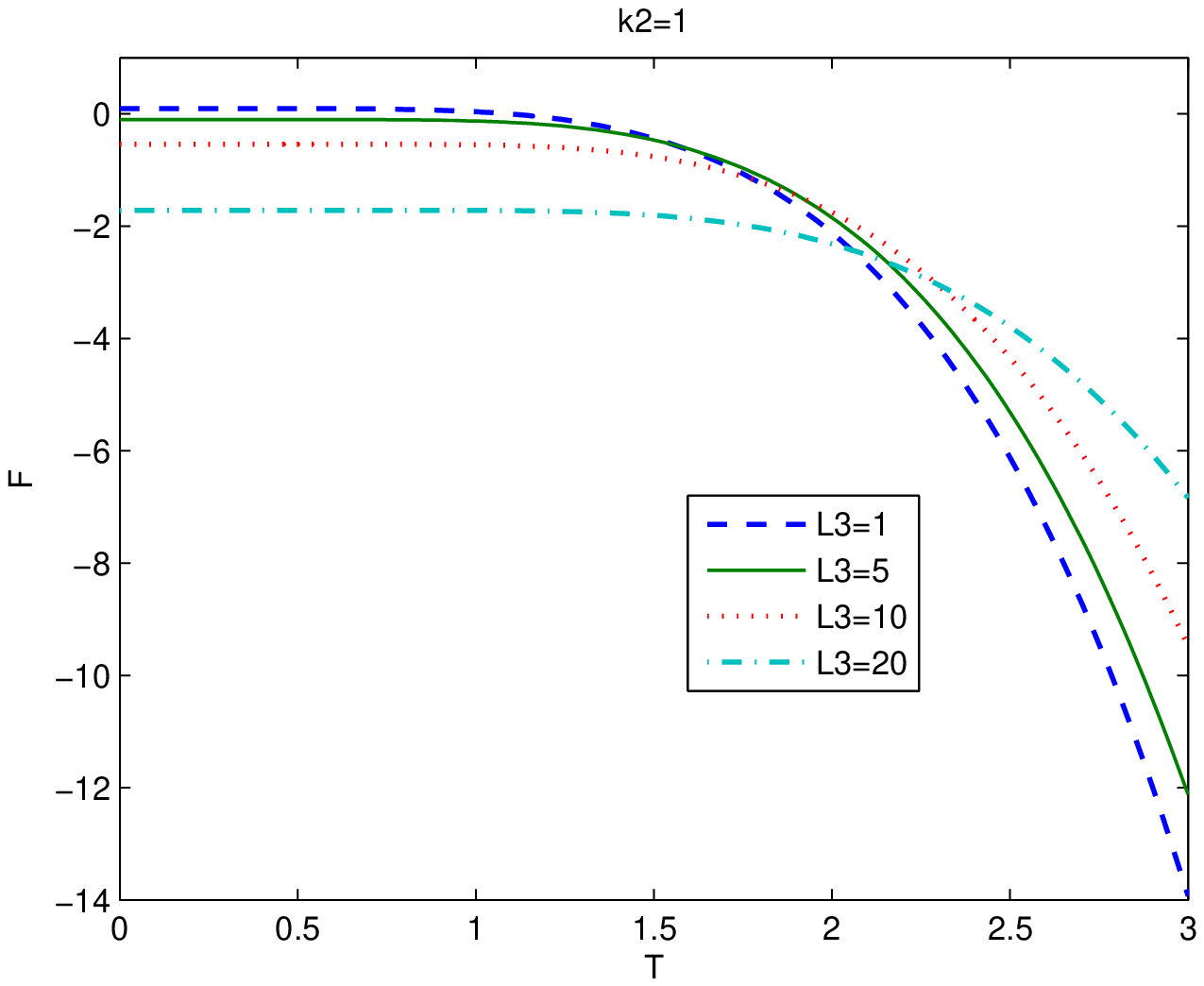}\epsfxsize=.45\linewidth
\epsffile{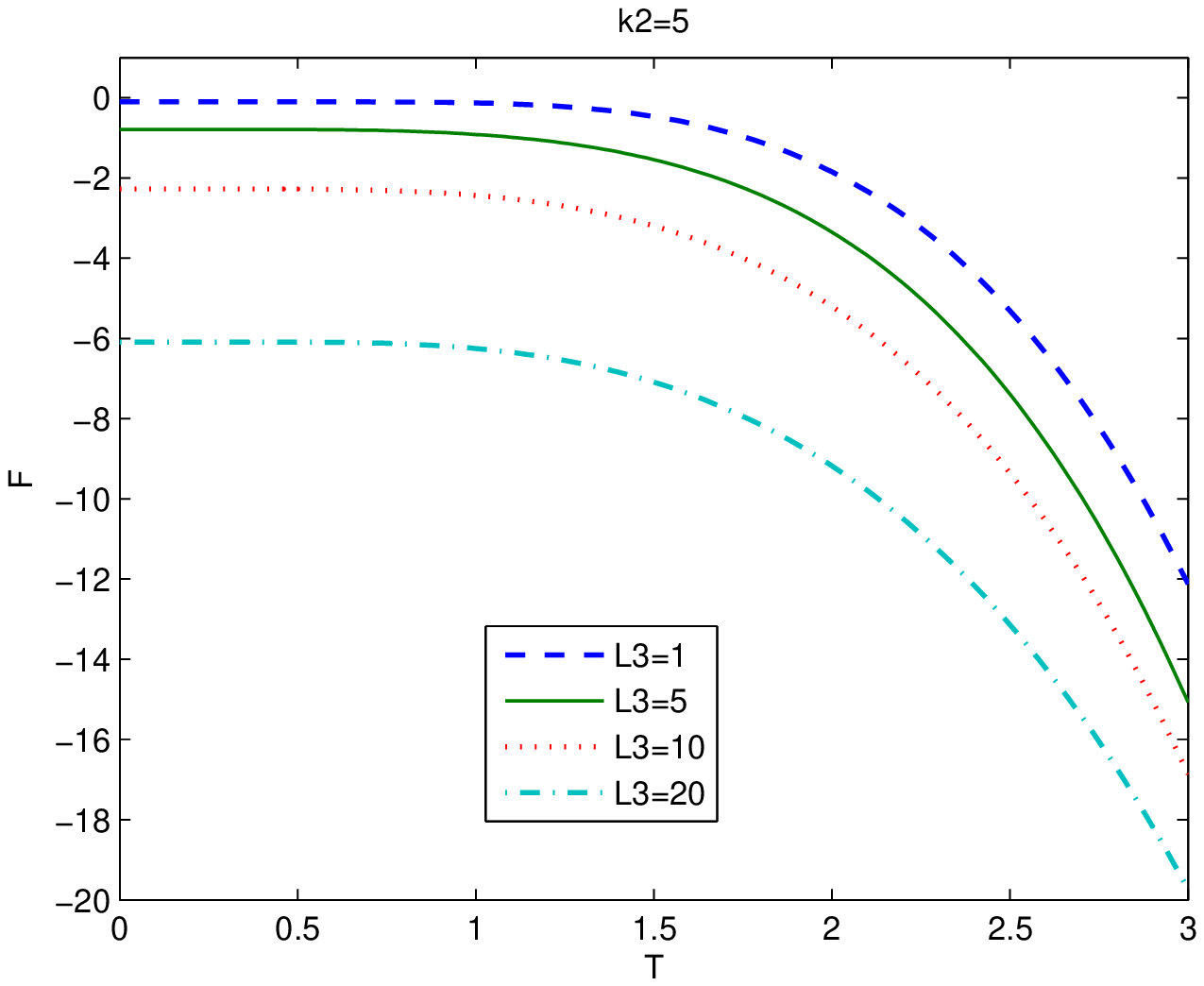}\caption{The free energy $F_{A_C}(L_1, L_2,
L_3)$ as a function of $T$ when $V=L_1L_2L_3=1$ and $k_2=L_2/L_1=1,
5$ at
 $k_3=L_3/L_1=1, 5, 10,20$ respectively. }\end{figure}

 Under the simultaneous space--time
scaling $\beta\mapsto \lambda \beta$, $L_i\mapsto \lambda L_i, 1\leq
i\leq d$, both $F_{A_C}( L_1, \ldots,  L_d)$ and $F_{A_B}(  L_1,
\ldots, L_d)$ transform as
\begin{align*}
F_{A_{C/B}}(  L_1, \ldots,  L_d)\mapsto \lambda^{-1}F_{A_{C/B}}(L_1,
\ldots, L_d),
\end{align*}and thus the thermodynamic relation \eqref{eq5_23_7}
holds.

The low and high temperature expansions of the free energy
$F_{A_{C/B}}$ can be obtained from the corresponding expansion of
$F_P$ using \eqref{eq27_3} and \eqref{eq27_3_1}. The sign of the
zero temperature energy $ F_{A_{C/B}}^0(L_1, \ldots, L_d)$ also
depends on the relative size of $L_1, \ldots, L_d$.

 \begin{figure}\centering \epsfxsize=.4\linewidth
\epsffile{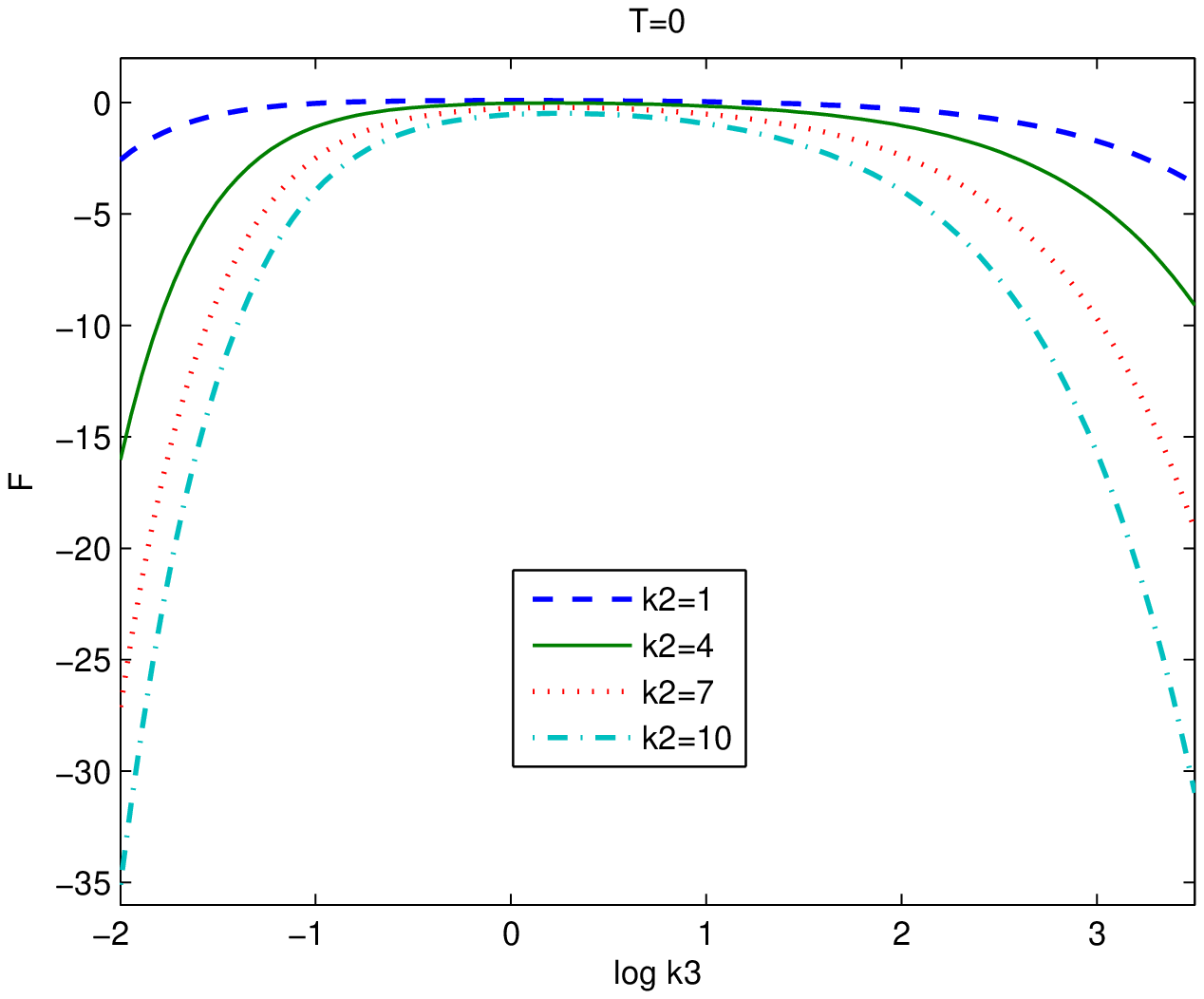}\epsfxsize=.4\linewidth
\epsffile{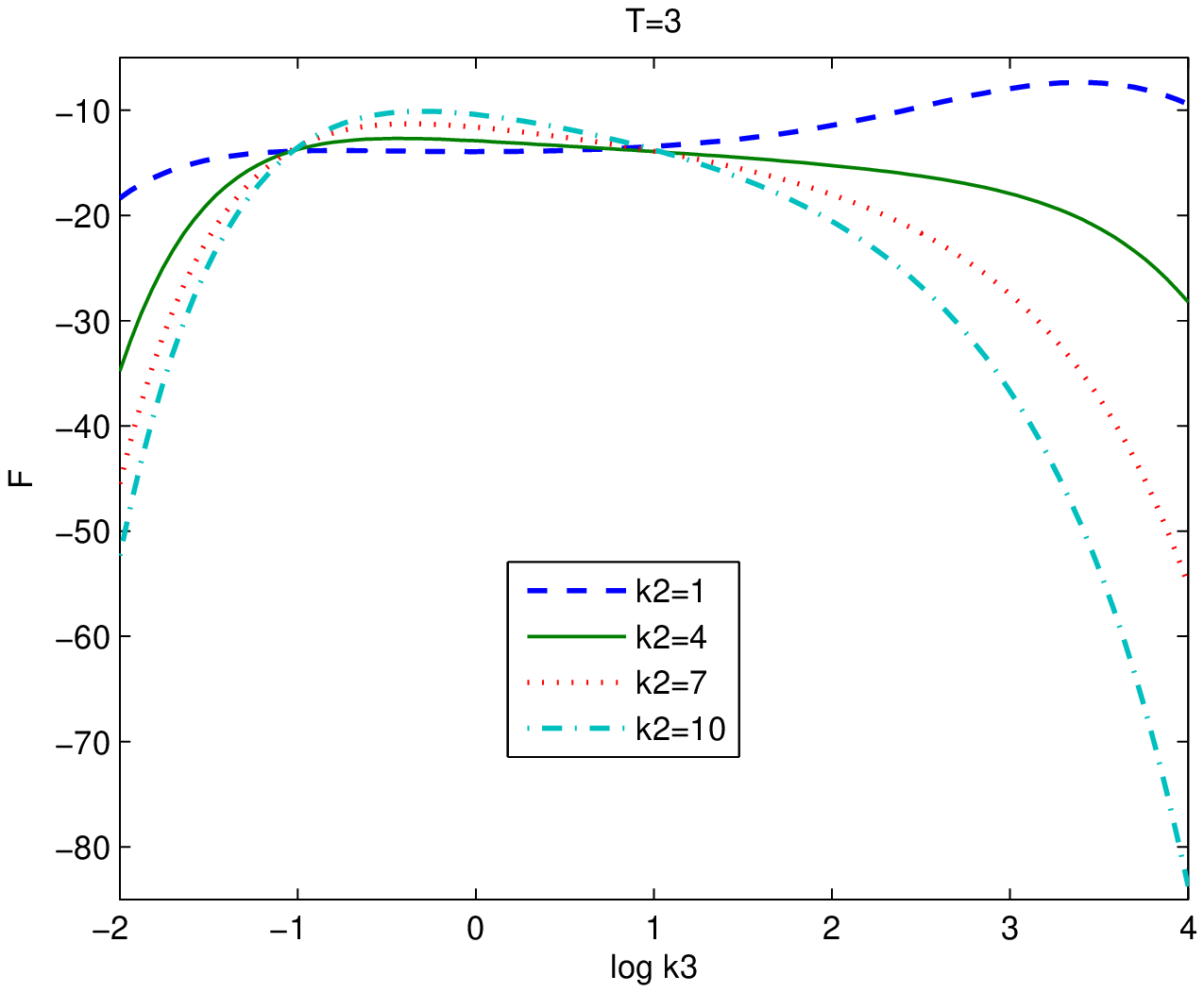}\\\epsfxsize=.4\linewidth
\epsffile{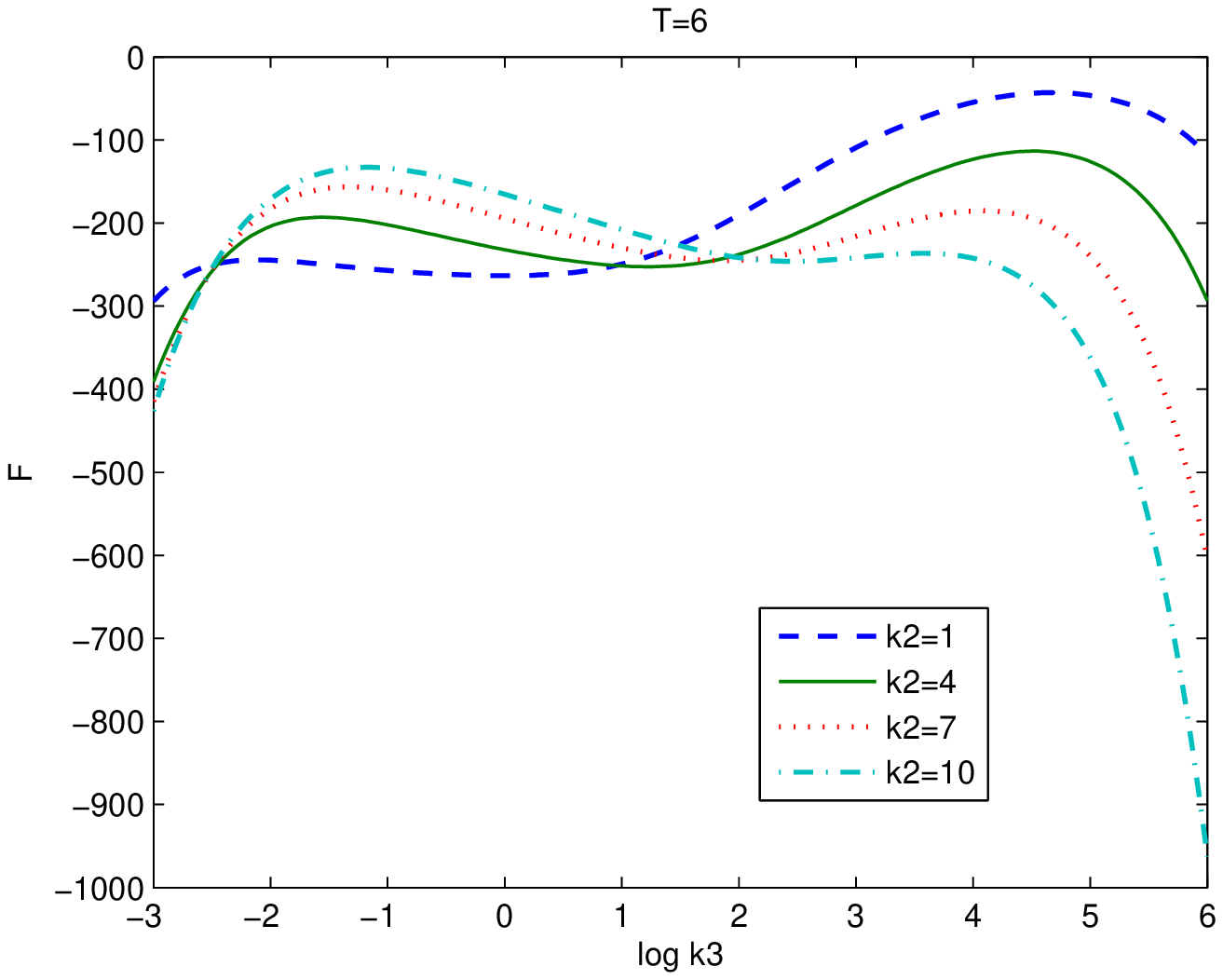}\epsfxsize=.4\linewidth
\epsffile{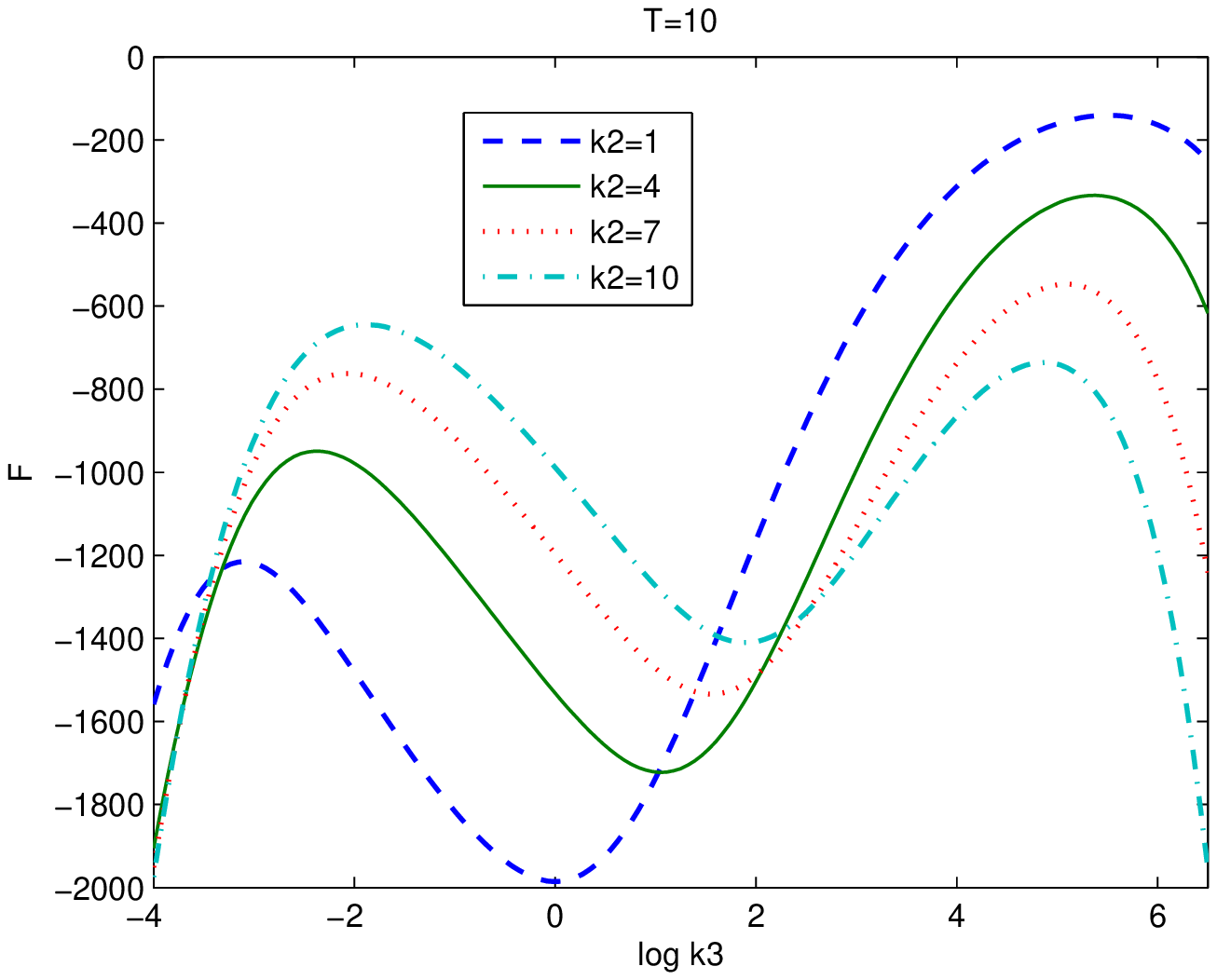}\caption{The free energy $F_{A_C}(L_1, L_2,
L_3)$ as a function of $k_3=L_3/L_1$ when  $k_2=L_2/L_1=1, 4, 7, 10$
and $V=L_1L_2L_3=1$, at $T=0,3,6,10$ respectively. }\end{figure}

\begin{figure}\centering \epsfxsize=.4\linewidth
\epsffile{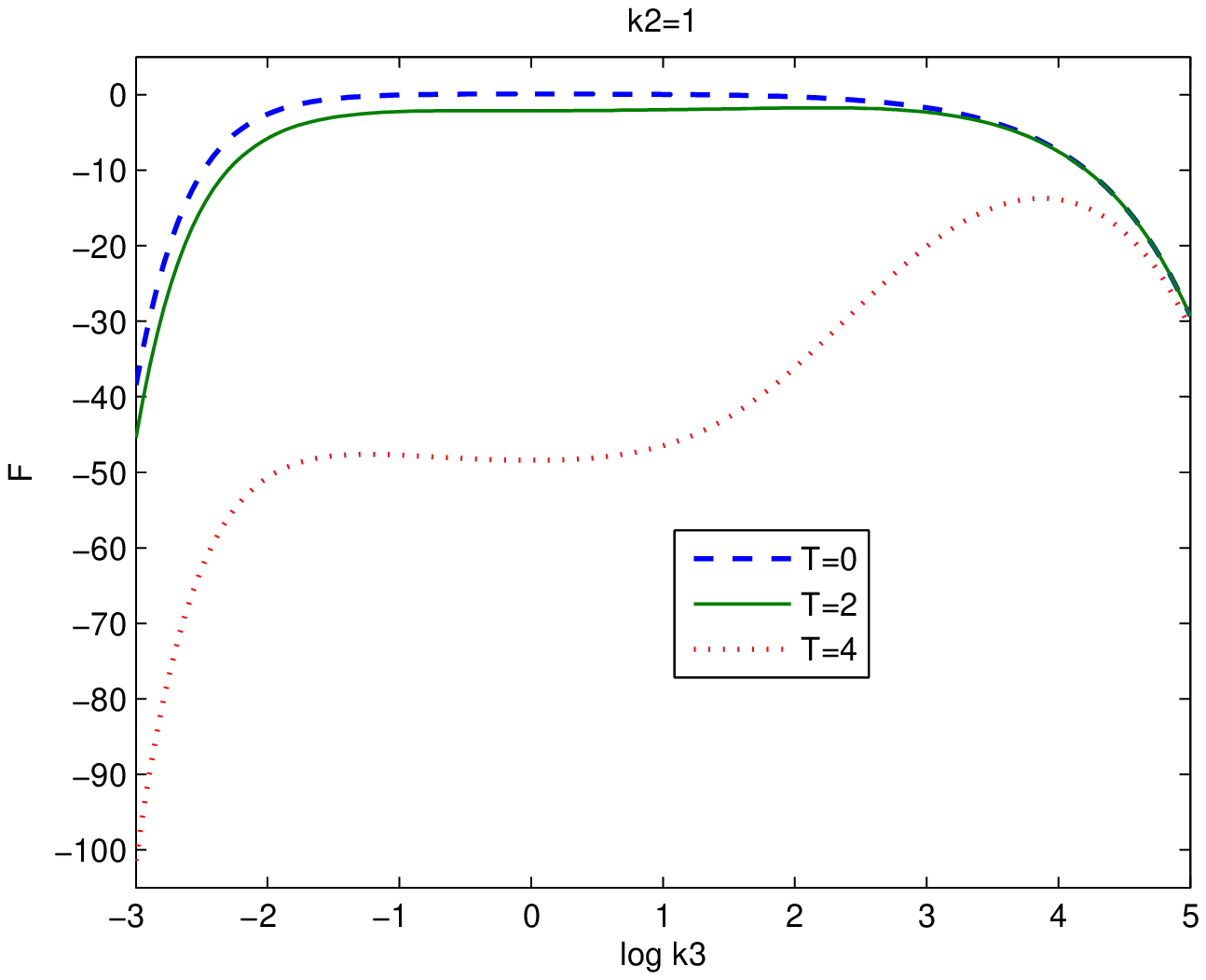}\epsfxsize=.4\linewidth
\epsffile{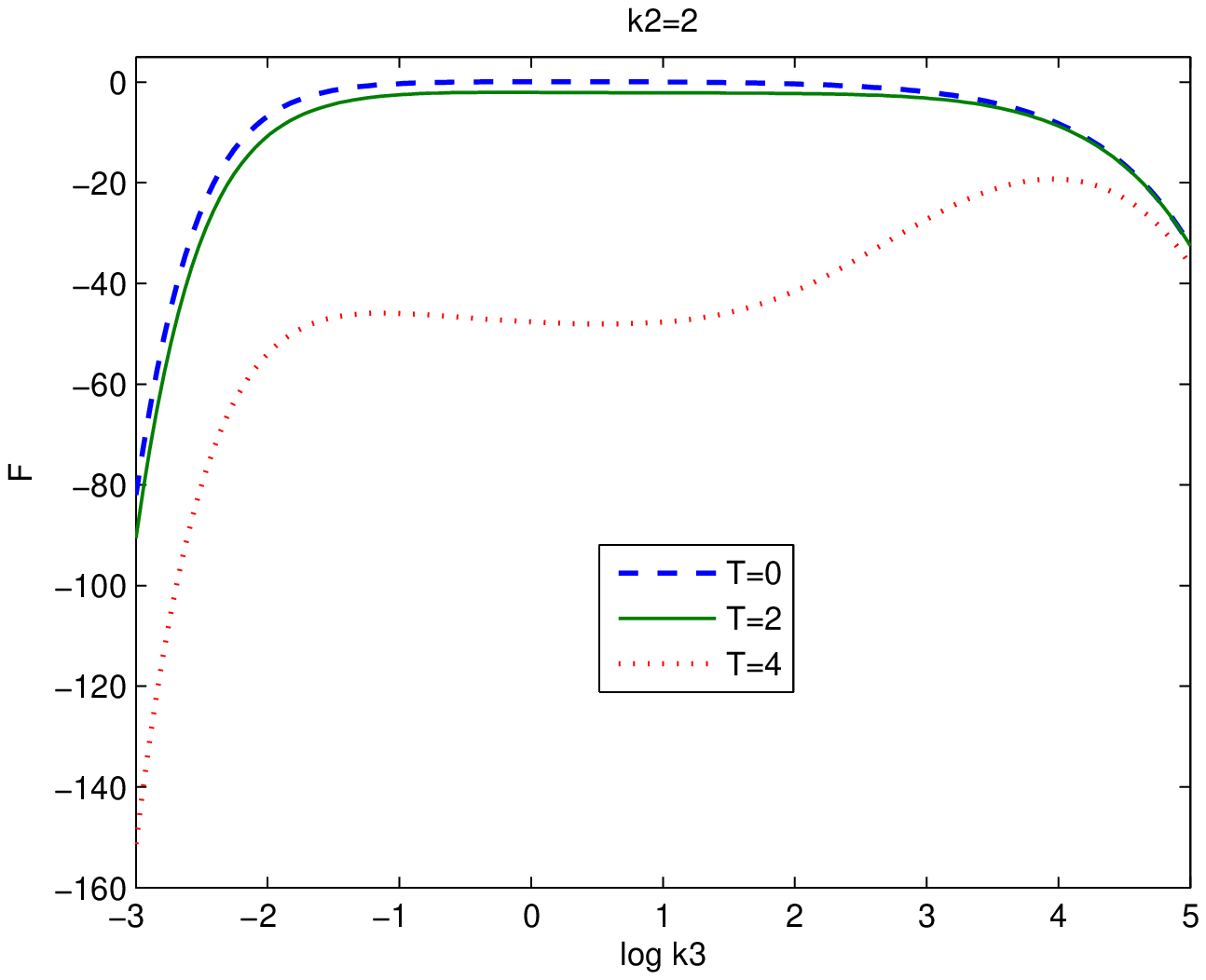}\\\epsfxsize=.4\linewidth
\epsffile{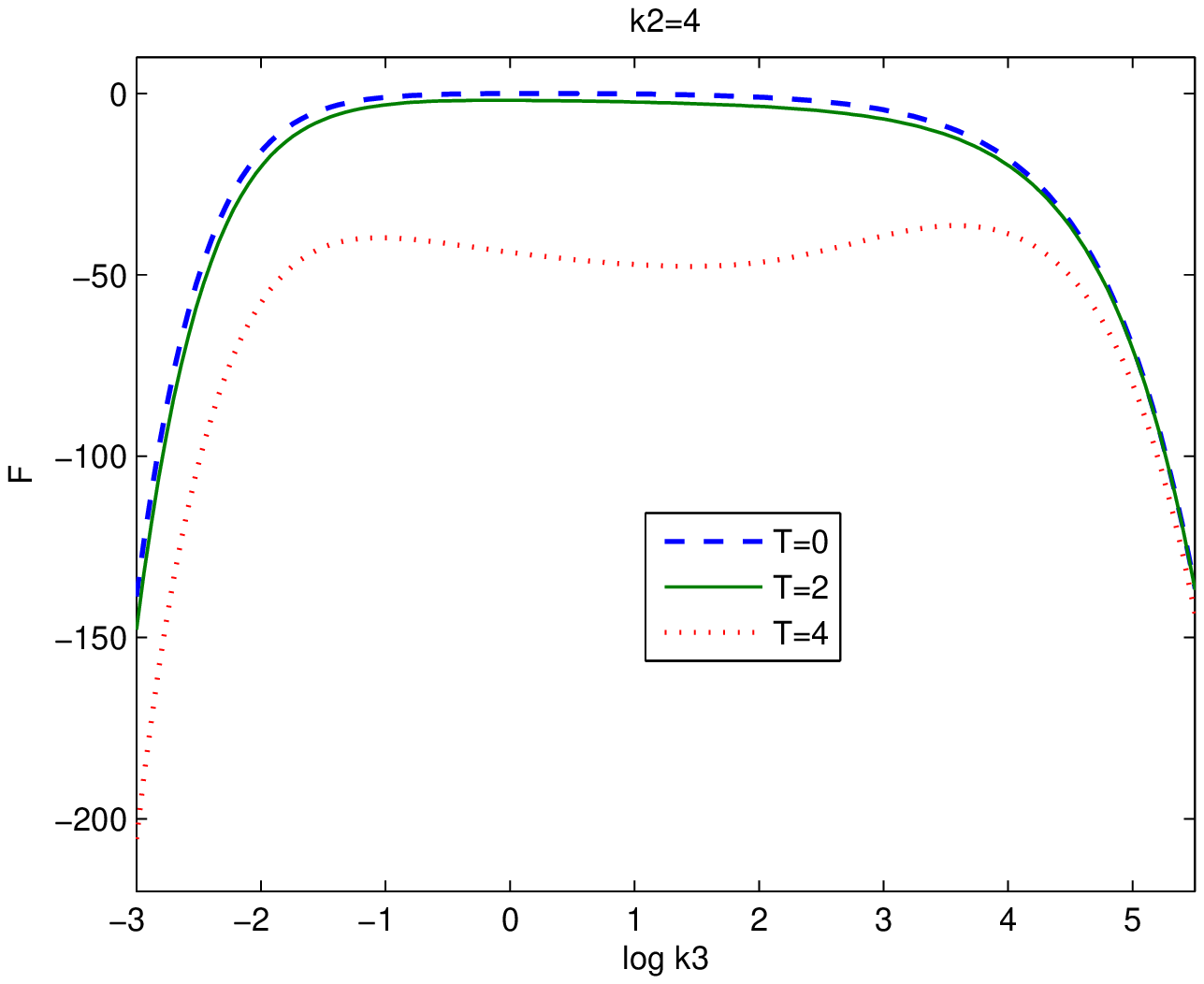}\epsfxsize=.4\linewidth
\epsffile{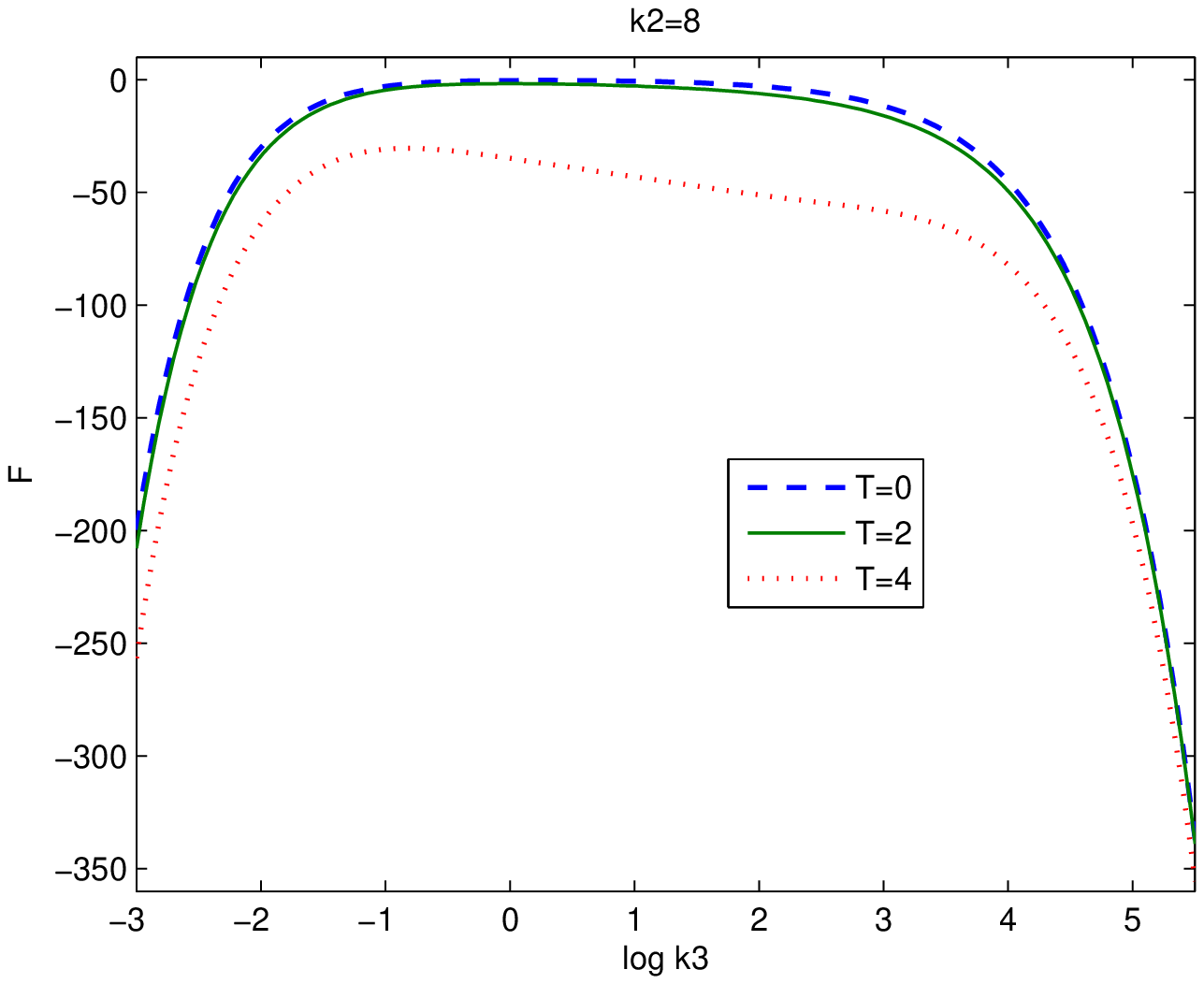}\caption{The free energy $F_{A_C}(L_1, L_2,
L_3)$ as a function of $k_3=L_3/L_1$ when  $T=0, 2, 4$  and
$V=L_1L_2L_3=1$ at $k_2=L_2/L_1=1,2,4,8$.}\end{figure}

\vspace{0.5cm}\vspace{0.1cm} \noindent\textbf{Table 2}\\\noindent
\begin{tabular}{c|c}
\hline
$k_2=L_2/L_1$ & The range of $k_3=L_3/L_1$ where $F_{A_C}^0(L_1, L_2, L_3)\geq 0$\\
\hline 0.75& $ 0.3555 \leq k_3\leq2.7033  $   \\
\hline
 1.& $0.4083 \leq k_3\leq 3.4298 $   \\ \hline
1.25& $0.4580\leq k_3\leq 3.6219$    \\
\hline 1.5 & $0.5057 \leq k_3\leq 3.4957$\\\hline
\end{tabular}\\
\\

\begin{figure}\centering
\epsfxsize=.45\linewidth
\epsffile{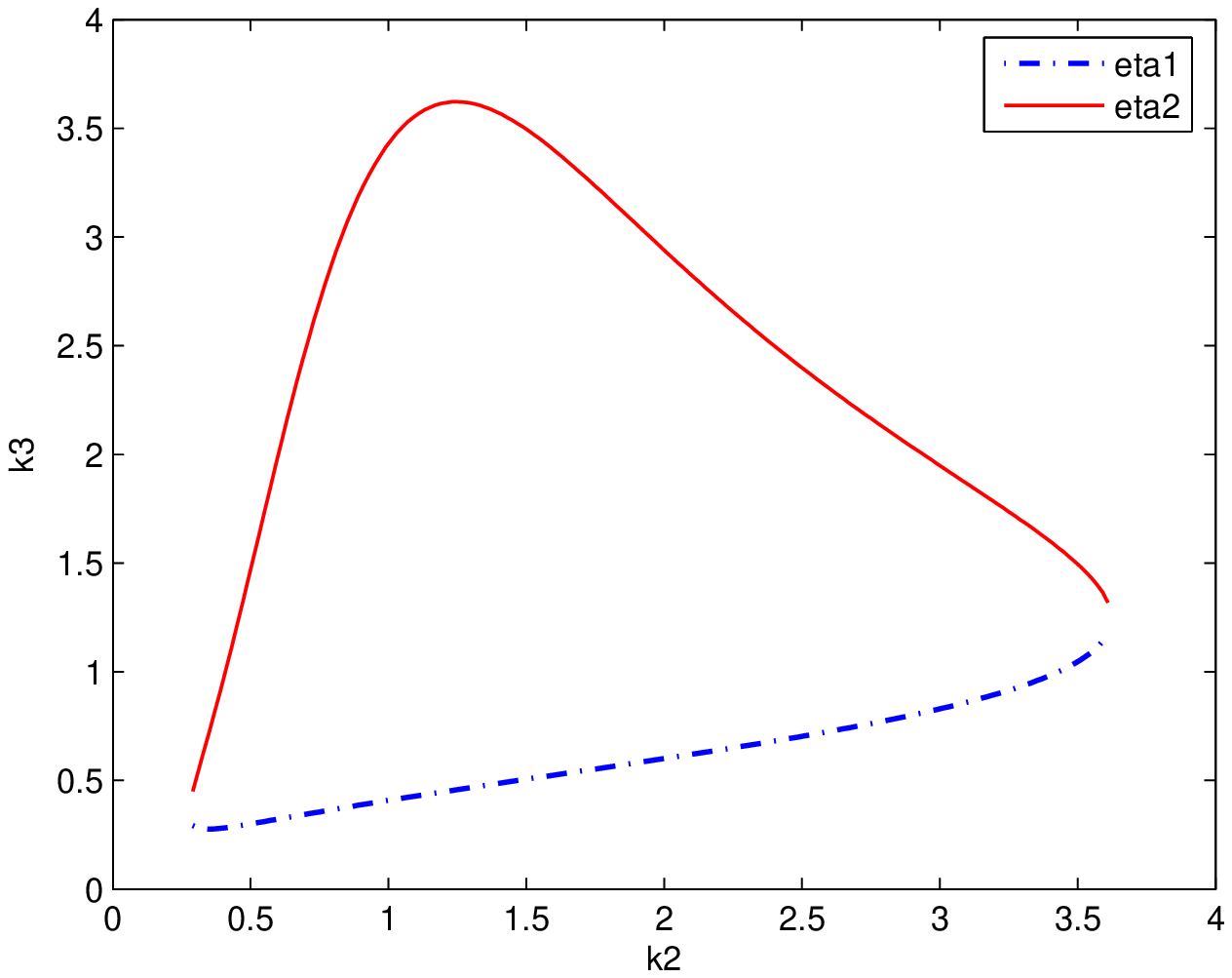}\epsfxsize=.45\linewidth
\epsffile{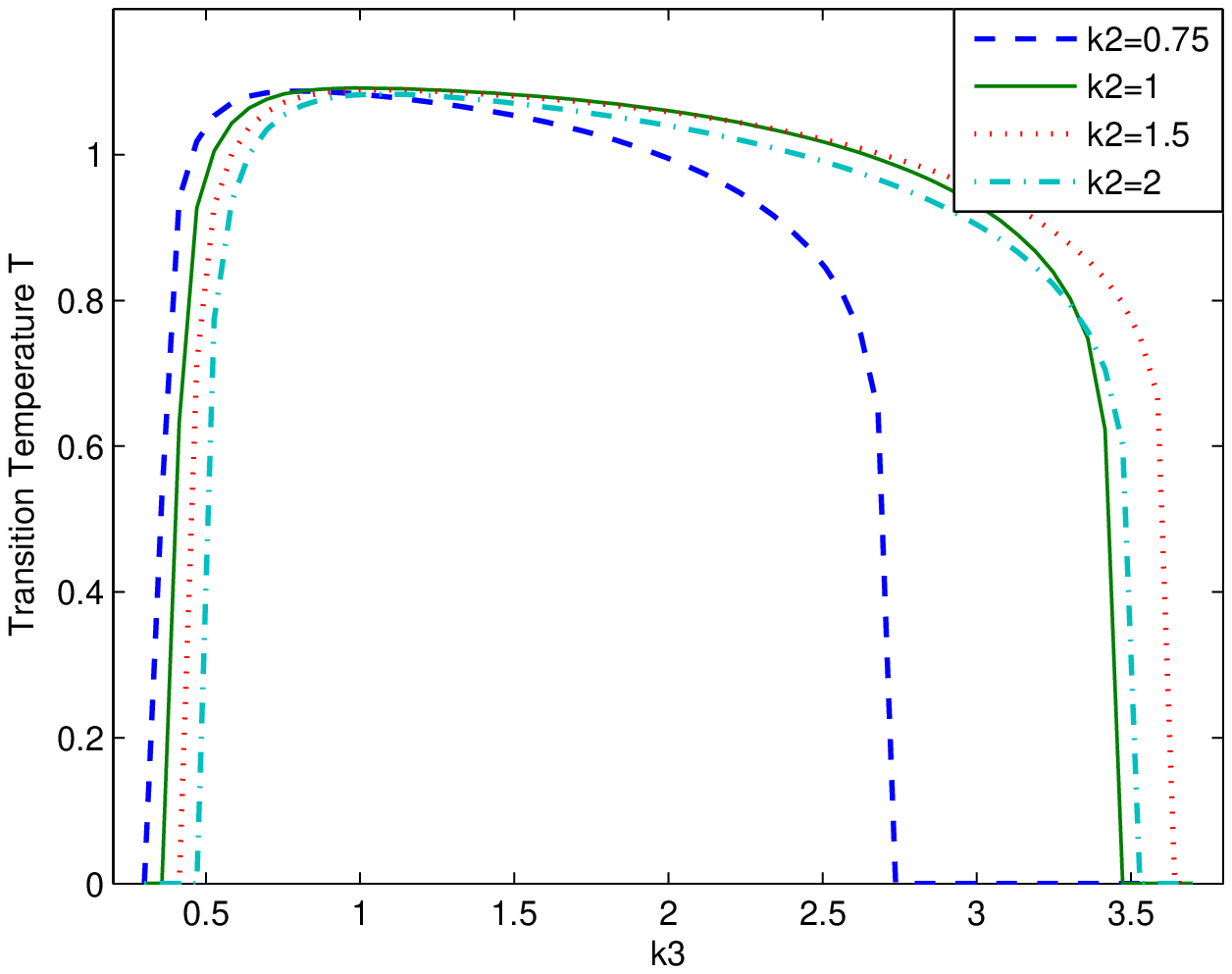}\caption{\textbf{Left:} When $0.2761\leq
k_2=L_2/L_1\leq 3.6195$, there exist $\eta_1(k_2)$ and $\eta_2(k_2)$
such that $F_{A_C}(L_1,L_2, L_3)\geq 0$ for all $\eta_1(k_2)\leq
L_3/L_1\leq \eta_2(k_2)$. The graph shows $\eta_1$ and $\eta_2$ as
functions of $k_2$. \textbf{Right:} The transition temperature
$T(L_1, L_2,L_3)$ for $F_{A_C}(L_1, L_2,L_3)$ as a function of
$k_3=L_3/L_1$ when $V=1$ and $k_2=1, 1.5, 2$. }
\end{figure}

In the high temperature regime,  the leading term is
$$-(d-1)\frac{L_1\ldots L_d}{\pi^{\frac{d+1}{2}}\beta^{d+1}
}\Gamma\left(\frac{d+1}{2}\right) \zeta_R(d+1),$$ which is $(d-1)$
times the leading term in the scalar field case. This is due to the
fact that electromagnetic field in $d+1$ dimensional space--time has
$d-1$ polarization states. The term proportional to
$\frac{1}{\beta}\log\beta$ still present. When $d$ is even, there
are terms proportional to $\frac{1}{\beta^j}$ for all $1\leq j\leq
d+1$. When $d$ is odd, there are terms proportional to
$\frac{1}{\beta^j}$ for all $1\leq j\leq d+1$ except for
$j=\frac{d+3}{2}$.

 When $d=3$, we find that the zero point energy
$F_{A_C}^0(L_1, L_2, L_3)$ is given by
\begin{align*} &F_{A_C}^0(L_1, L_2, L_3)=F_{A_B}^0(L_1, L_2, L_3)\\=&
-\frac{L_1L_2L_3}{16\pi^2}\sum_{\mathbf{k}\in
\Z^3\setminus\{\mathbf{0}\}}\frac{1}{\left((L_1k_1)^2+(L_2k_2)^2+(L_3k^3)^2\right)^2}
+\frac{\pi}{48
}\left(\frac{1}{L_1}+\frac{1}{L_2}+\frac{1}{L_3}\right),\end{align*}
which is a well known result (see e.g., \cite{Mac}). We show
graphically some particular values of the transition temperature
$T(L_1,L_2,L_d)$ for $F_{A_C}(L_1,L_2, L_3)$ in Figure 10.

On the other hand, the high temperature expansion of $F_{A_C}(L_1,
L_2, L_3)=F_{A_B}(L_1, L_2, L_3)$ is
\begin{align}\label{eq5_23_6}
&F_{A_C}(L_1, L_2, L_3)=F_{A_B}(L_1, L_2,
L_3)=-\frac{\pi^2}{45}\frac{L_1L_2L_3}{\beta^4}+\frac{\pi}{12}\frac{L_1+L_2+L_3}{\beta^2}\\&
+\frac{1}{2\beta}\log\beta-\frac{1}{8\beta}Z_{E,3}'\left(0;L_1^{-1},
L_2^{-1}, L_3^{-1}\right)-\frac{1}{4\beta}\log (8\pi L_1 L_2 L_3
)\nonumber\\&-\frac{L_1L_2L_3}{2\beta^2}\sum_{\mathbf{k}\in\Z^{3}\setminus\{\mathbf{0}\}}
\frac{1}{\sum_{j=1}^3[L_jk_j]^2}\frac{e^{\frac{4\pi
}{\beta}\sqrt{\sum_{j=1}^3[L_jk_j]^2}}}{\left(e^{\frac{4\pi
}{\beta}\sqrt{\sum_{j=1}^3[L_jk_j]^2}}-1\right)^2}\nonumber\\&
-\frac{L_1L_2L_3}{8\pi\beta}\sum_{\mathbf{k}\in\Z^{3}\setminus\{\mathbf{0}\}}
\frac{1}{\left[\sum_{j=1}^3[L_jk_j]^2\right]^{3/2}} \frac{1}{
e^{\frac{4\pi
}{\beta}\sqrt{\sum_{j=1}^3[L_jk_j]^2}}-1}\nonumber\\
&+\frac{1}{2\beta}\sum_{k=1}^{\infty}\frac{1}{k}\left[\frac{1}{e^{\frac{4\pi
k L_1 }{\beta}}-1}+\frac{1}{e^{\frac{4\pi  k
L_2}{\beta}}-1}+\frac{1}{e^{\frac{4\pi k L_3
}{\beta}}-1}\right].\nonumber
\end{align}Our result gives the correct  high
temperature limit stipulated by Ambj{\o}rn and Wolfram \cite{AW}
(equation (7.12)). However, they only obtained the first three
terms.  To the best of our knowledge, we are not aware of any
existing study that calculate the high temperature limit to the
degree of accuracy obtained here. We would like to emphasize that
the formula \eqref{eq5_23_6} is valid for all temperature. In
\cite{JVH}, the authors calculate this free energy by a different
method. They gave the same first two leading terms as above, and no
explicit formulas for the remaining terms are given.

\section{From closed cavity to general case}

There exist many papers on  the Casimir energy of massless scalar
field or Casimir energy of electromagnetic field confined in a
$p$-dimensional rectangular cavity in a $d$-dimensional space
\cite{AW, OS, Ch1, CNS, Zheng2, Actor2, Zheng, Zh1, Inui1, Inui2,
Ed1, Li}. A $p$-dimensional rectangular cavity in a $d$-dimensional
space is a space of the form $\Omega_{p,d}=[0,
L_1]\times\ldots\times[0, L_p]\times \R^{d-p}$, where $0\leq p\leq
d$.  It can be considered as the limiting case of the closed cavity
where $L_1, \ldots, L_p \ll L_{p+1}=\ldots=L_d=L$ or
$L_{j}\rightarrow \infty$ for $p+1\leq j\leq d$. In  the existing
literature, when calculating the Casimir free energy, usually after
setting up the zeta function over $(k_1, \ldots, k_d)$ in a suitable
set, the summation over $k_{p+1}, \ldots, k_d$ is changed to
integration. From the mathematical point of view, this is not a
rigorous treatment since the summation expression for the zeta
function only works for $\text{Re}\; s>\frac{d}{2}$, which does not
include the point $s=0$. To justify this procedure, one actually
need to justify that the processes of taking analytic continuation
and taking limit $L_{j}\rightarrow \infty$ for $p+1\leq j\leq d$ can
be interchanged. In this section, we will directly take the limit
$L_{j}\rightarrow \infty$ for $p+1\leq j\leq d$ in the expression of
Casimir energy for fields inside a closed rectangular cavity to
obtain the energy of the fields inside a non-closed rectangular
cavity. To be more precise, the limit when $L_{j}\rightarrow \infty$
for $p+1\leq j\leq d$ of the free energy $F(L_1,\ldots, L_d)$ is
always infinite. Therefore we shall consider the free energy density
$f$ defined as the limit
\begin{align*}
f_d(L_1,\ldots, L_p)=\lim_{\substack{L_i\rightarrow \infty\\ p+1\leq
i\leq d}}\frac{F(L_1,\ldots, L_d)}{L_{p+1}\ldots L_d}.
\end{align*}

In the following, we assume that $0\leq p\leq d-1$. By putting
$m=d-p$, $a_j=2\pi/L_{p+j}, 1\leq j\leq d-p$, $a_{d-p+1}=2\pi/\beta$
and $a_j =2\pi/L_{j-d+p-1}, d-p+2\leq j\leq d+1$ in \eqref{eq25_10},
we find that the free energy $F_{P}(L_1, \ldots, L_d)$
\eqref{eq30_3} is equal to
\begin{align*}
F_P(L_1,\ldots, L_d)=&-\frac{1}{2\beta}Z_{E,d-p}'\left(0;
\frac{2\pi}{L_{p+1}},\ldots,
\frac{2\pi}{L_d}\right)-\frac{1}{\beta}\log\beta\\&-\frac{\pi^{-\frac{d+1}{2}}\Gamma\left(\frac{d+1}{2}\right)}{2(2\pi)^{d+1}}L_1\ldots
L_{d} Z_{E, p+1}\left(\frac{d+1}{2};
\frac{\beta}{2\pi},\frac{L_1}{2\pi}, \ldots,
\frac{L_p}{2\pi}\right)\\&-\frac{1}{2\beta}R_{n,d-p}\left(\frac{2\pi}{\beta},
\frac{2\pi}{L_1}, \ldots, \frac{2\pi}{L_d}\right),
\end{align*}where $R_{n,m}(a_1, \ldots, a_n)$ is defined by \eqref{eq30_4}.
Now the last term goes to zero as $L_{j}\rightarrow \infty$ for
$p+1\leq j\leq d$ (see appendix). Therefore,
\begin{align}\label{eq13_1}
f_{P,d}(L_1,\ldots,
L_p)=&-\frac{1}{2\beta}\lim_{\substack{L_i\rightarrow
\infty\\p+1\leq i\leq d}}\frac{1}{ L_{p+1}\ldots,
L_{d}}Z_{E,d-p}'\left(0; \frac{2\pi}{L_{p+1}},\ldots,
\frac{2\pi}{L_d}\right)\\&-\frac{\pi^{-\frac{d+1}{2}}\Gamma\left(\frac{d+1}{2}\right)}{2(2\pi)^{d+1}}L_1\ldots
L_{p} Z_{E, p+1}\left(\frac{d+1}{2};
\frac{\beta}{2\pi},\frac{L_1}{2\pi}, \ldots,
\frac{L_p}{2\pi}\right).\nonumber
\end{align}
Next we want to show that the first term in \eqref{eq13_1}  is also
zero, i.e. we need to show that
$$\lim_{\substack{a_i\rightarrow 0\\1\leq i\leq n}}a_1\ldots
a_{n}Z_{E,n}'\left(0; a_1,\ldots, a_n\right)=0.$$ When $n=1$, we
have
\begin{align*}
Z_{E,1}'(0; a) =2\log\frac{a}{2\pi},
\end{align*}therefore $\lim_{a\rightarrow 0} \left[aZ_{E,1}'(0; a)\right] =0$.
When $n>1$,  equation \eqref{eq25_10}   gives
\begin{align*}
&Z_{E,n}'\left(0; a_1,\ldots, a_n\right)\\=&Z_{E,n-1}'\left(0;
a_1,\ldots, a_{n-1}\right)
+\frac{2\pi^{-n/2}a_n^{n-1}\Gamma\left(\frac{n}{2}\right)}{\left[\prod_{j=1}^{n-1}a_j\right]}
\zeta_R(n)+R_{n,n-1}( a_1, \ldots, a_n).
\end{align*}Using a similar argument as before (see appendix),
\begin{align*}
\lim_{\substack{a_i\rightarrow 0\\1\leq i\leq n-1}}a_1\ldots
a_{n-1}R_{n,n-1}( a_1, \ldots, a_n)=0.
\end{align*}On the other hand, it is obvious that
\begin{align*}
\lim_{\substack{a_i\rightarrow 0\\1\leq i\leq n-1}}a_1\ldots
a_{n}\frac{2\pi^{-n/2}a_n^{n-1}\Gamma\left(\frac{n}{2}\right)}{\left[\prod_{j=1}^{n-1}a_j\right]}
\zeta_R(n)=0.
\end{align*}Therefore,
\begin{align*}
\lim_{\substack{a_i\rightarrow 0\\1\leq i\leq n}}a_1\ldots
a_{n}Z_{E,n}'\left(0; a_1,\ldots,
a_n\right)=\lim_{\substack{a_i\rightarrow 0\\1\leq i\leq
n}}a_1\ldots a_{n}Z_{E,{n-1}}'\left(0; a_1,\ldots, a_{n-1}\right),
\end{align*}and we obtain by induction on $n$ that this is zero.
Consequently, we find from \eqref{eq13_1} that
\begin{align}\label{eq4_5}
f_{P,d}(L_1,\ldots,
L_p)=-\frac{\Gamma\left(\frac{d+1}{2}\right)}{2\pi^{\frac{d+1}{2}}
}L_1\ldots L_{p} Z_{E, p+1}\left(\frac{d+1}{2}; \beta,L_1, \ldots,
L_p\right)
\end{align}and this agrees with the result in \cite{AW} obtained by dimensional regularization method.
 Notice that the right hand side of \eqref{eq4_5} is not defined
when $p=d$. Under the simultaneous space--time scaling
$\beta\rightarrow \lambda\beta$, $L_i\mapsto \lambda L_i, 1\leq
i\leq d$, the free energy density $f_{P,d}(L_1,\ldots, L_p)$
transforms as \begin{align}\label{eq11_1} f_{P,d}(L_1,\ldots,
L_p)\mapsto \lambda^{p-d-1}f_{P,d}(L_1,\ldots, L_p).
\end{align}
Now using the fact that
\begin{align*}& \lim_{\substack{L_j\rightarrow \infty\\ p+1\leq j\leq
d}}\frac{F_P\left(2L_{m_1},\ldots,2L_{m_j}, 2L_{p+1},\ldots
,2L_d\right)}{L_{p+1} \ldots L_d}\\=& 2^{d-p}
\lim_{\substack{L_j\rightarrow \infty\\ p+1\leq j\leq
d}}\frac{F_P\left(2L_{m_1},\ldots,2L_{m_j}, 2L_{p+1},\ldots
,2L_d\right)}{(2L_{p+1}) \ldots
(2L_d)}=2^{d-p}f_{P,d}\left(2L_{m_1},\ldots, 2L_{m_j}\right),
\end{align*}
and \eqref{eq3_1}, \eqref{eq27_3}, we find that the free energy
densities $f_{D,d}(L_1, \ldots, L_p)$, $f_{N,d}(L_1,\ldots, L_p)$,
$f_{A_C,d}(L_1,\ldots, L_p)$, $f_{A_B,d}(L_1,\ldots, L_p)$ for
massless scalar field under Dirichlet and Neumann boundary
conditions and for electromagnetic field confined in cavity with
perfectly conducting walls and with infinitely permeable walls are
related to the free energy for massless scalar field under periodic
condition by:
\begin{align}\label{eq4_7}
&f_{D/N,d}(L_1,\ldots,L_p)=2^{-p}\sum_{j=0}^{p}(\mp 1)^{p-j}
\sum_{1\leq m_1<\ldots<m_j\leq p}f_{P,j+d-p}(2L_{m_1}, \ldots,
2L_{m_j})\\
&f_{A_{C/B,d}}(L_1,\ldots,L_p)\nonumber\\=&2^{-p}\sum_{j=0}^{p}(\mp
1)^{p-j}(d-1-2p+2j) \sum_{1\leq m_1<\ldots<m_j\leq
p}f_{P,j+d-p}(2L_{m_1}, \ldots, 2L_{m_j}).\nonumber
\end{align}
The scaling behavior of the  free energy density in these cases is
the same as the periodic case \eqref{eq11_1}.

 When
$p=0$, we obtain the vacuum energy of free massless scalar field and
electromagnetic field in $\R^d$:
\begin{align}\label{eq9_1}
f_{P,d}=f_{D,d}&=f_{N,d}=-\frac{\Gamma\left(\frac{d+1}{2}\right)}{\pi^{\frac{d+1}{2}}}\frac{1}{\beta^{d+1}}\zeta_{R}(d+1),\\
f_{A_{C/B,d}}=&-(d-1)\frac{\Gamma\left(\frac{d+1}{2}\right)}{\pi^{\frac{d+1}{2}}}\frac{1}{\beta^{d+1}}\zeta_{R}(d+1),\nonumber
\end{align}which are the Stefan-Boltzmann terms. These   equations
are reasonable since when extending to the space $\R^d$, the
boundary disappears and the vacuum energy should be the same no
matter what boundary conditions we start with.

\vspace{0.2cm}
\subsection{Low Temperature Expansion}~

\vspace{0.2cm}\noindent When $1\leq p\leq d-1$, by putting $m=p$,
$a_i=L_i, 1\leq i\leq p$, $a_{p+1}=\beta$ in the Chowla-Selberg
formula \eqref{eq31}, we obtain the low temperature ($T\ll 1$)
expansion of the free energy density \eqref{eq4_5}:
\begin{align}\label{eq4_8}
&f_{P,d}(L_1,\ldots,
L_p)\\=&-\frac{\Gamma\left(\frac{d+1}{2}\right)}{2\pi^{\frac{d+1}{2}}}L_1\ldots
L_pZ_{E,p}\left(\frac{d+1}{2}; L_1,\ldots, L_p\right)
-\frac{\Gamma\left(\frac{d-p+1}{2}\right)}{\pi^{\frac{d-p+1}{2}}}\frac{\zeta_R(d-p+1)
}{\beta^{d-p+1}}\nonumber\\
&-\frac{2}{\beta^{\frac{d-p+1}{2}}}\sum_{m=1}^{\infty}\sum_{\mathbf{k}\in
\Z^{p}\setminus\{\mathbf{0}\}}\frac{1}{m^{\frac{d-p+1}{2}}}\left(\sum_{j=1}^p
\left[\frac{k_j}{L_j}\right]^2\right)^{\frac{d+1-p}{4}}K_{\frac{d-p+1}{2}}\left(2\pi
m\beta \sqrt{\sum_{j=1}^p
\left[\frac{k_j}{L_j}\right]^2}\right).\nonumber
\end{align}
The first term gives the zero temperature energy density and the sum
of the last two terms is the thermal correction. Note that now the
thermal correction contains a term proportional to
$\beta^{-\frac{d-p+1}{2}}$. As usual the last term decays
exponentially. We show in the appendix that the sum of the  thermal
correction is equal to
\begin{align*}
\frac{1}{(2\pi)^{d-p}\beta}\int_{\R^{d-p}}\sum_{\mathbf{k}\in
\Z^{p}}\log\left(1-e^{-\beta\sqrt{\sum_{j=1}^p\left[\frac{2\pi
k_j}{L_j}\right]^2+ |\mathbf{w}|^2}}\right)dw_1\ldots dw_{d-p},
\end{align*}in agreement with the usual integration prescription to obtain the limit $L_j\rightarrow \infty$
for $p+1\leq j\leq d$. From this formula, we can verify as in the
closed cavity case that the free energy density is a decreasing
function of temperature. On the other hand, \eqref{eq4_5} implies
that  the Casimir free energy is negative at all temperature for all
$p$ and $d$ such that $0\leq p< d$.

Compare to \eqref{eq56}, we find that we cannot simply set $p=d$ in
\eqref{eq4_8} to obtain the free energy in the closed cavity case
\eqref{eq56} due to the second term. In fact, by using physics
argument, Ambj{\o}rn and Wolfram \cite{AW} has argued that in order
to obtain the free energy for closed cavity from this formula, it is
necessary to omit the second term.

Using \eqref{eq4_7} and \eqref{eq4_8}, one can also obtain the low
temperature expansion of the free energy densities $f_{D/N,d}$ and
$f_{A_{C/B,d}}$ for $1\leq p\leq d-1$.
 We find that in the case of scalar field with
Dirichlet boundary condition, the thermal correction is an
exponentially decay term, whereas for the scalar field with Neumann
boundary condition and also for electromagnetic field confined in a
cavity with infinitely permeable walls, there is an extra term
proportional to $\beta^{-\frac{d-p+1}{2}}$ and for the
electromagnetic field confined in a cavity with perfectly conducting
walls, this extra term only present when $p=1$. Just like the
periodic case, we can show that the thermal corrections are equal to
\begin{align}\label{eq10_4}
\frac{1}{(2\pi)^{d-p}\beta}\int_{\R^{d-p}}\sum_{\mathbf{k}\in
(\mathbb{N}\cup\{0\})^p}M_{\mathcal{BC}}(\mathbf{k})\log\left(1-e^{-\beta\sqrt{\sum_{j=1}^p\left[\frac{\pi
k_j}{L_j}\right]^2+ |\mathbf{w}|^2}}\right)dw_1\ldots dw_{d-p},
\end{align}which is in agreement with the usual integration prescription.
Here $\mathcal{BC}=D,N, A_C, A_B$ and \begin{align*}
M_{D}(\mathbf{k})=&\begin{cases} 1, \hspace{1cm}&\text{if}\;\;
\mathbf{k}\in\mathbb{N}^p,\\
0, &\text{otherwise},\end{cases}
\\M_{N}(\mathbf{k})=&\;1\hspace{1cm}\forall \mathbf{k}\in
(\mathbb{N}\cup\{0\})^p,\\
M_{A_C}(\mathbf{k})=&\begin{cases} d-1,\hspace{1cm}&\text{if}\;\;k_i\neq 0\; \text{for all}\; 1\leq i\leq p,\\
1, &\text{if}\;\; k_i=0 \;\text{for some $i$, and $k_j\neq 0$ for
all other $j\neq i$},\\
0, &\text{otherwise}.
\end{cases}\\M_{A_B}(\mathbf{k})=& \; d-p+j-1, \hspace{1cm} \text{if exactly  $j$ of the $k_1, \ldots, k_p$ are nonzero.}
\end{align*} From this, we can also conclude that the free energy density is a
decreasing function of temperature. In the case of scalar field with
Neumann condition, we can even generalize the conclusion to that the
Casimir energy is always negative. However, in the case of scalar
field with Dirichlet condition and the cases of electromagnetic
fields, the sign of the Casimir free energy depends on $p, d, T$ and
the values of $L_1,\ldots, L_p$. There have been some discussions on
this point in \cite{Ch1, CNS, Zh1, Li}.


\vspace{0.2cm}
\subsection{High Temperature Expansion}~

\vspace{0.2cm} \noindent When $1\leq p\leq d-1$, by putting $m=1$,
$a_1=\beta$, $a_{i}=L_{i-1}, 2\leq i\leq p+1$, in the Chowla-Selberg
formula \eqref{eq31}, we obtain the high temperature ($T\gg 1$)
expansion of the free energy density \eqref{eq4_5}:
\begin{align}\label{eq9_2}
&f_{P,d}(L_1, \ldots,
L_p)=-\frac{\Gamma\left(\frac{d+1}{2}\right)}{\pi^{\frac{d+1}{2}}}L_1\ldots
L_p \frac{\zeta_R(d+1)}{\beta^{d+1} }\\\nonumber
&-\frac{\Gamma\left(\frac{d}{2}\right)}{2\pi^{\frac{d}{2}}}L_1\ldots
L_p\frac{1}{\beta}Z_{E,p}\left(\frac{d}{2}; L_1, \ldots,
L_p\right)\\&-\frac{2L_1\ldots L_p}{\beta^{\frac{d+2}{2}}}
\sum_{\mathbf{k}\in\Z^{p}\setminus\{\mathbf{0}\}}
\sum_{m=1}^{\infty}m^{\frac{d}{2}}\left(\sum_{j=1}^{p}\left[L_jk_j\right]^2\right)^{-\frac{d}{4}}
K_{\frac{d}{2}}\left(\frac{2\pi m}{\beta}
\sqrt{\sum_{j=1}^{p}\left[L_jk_j\right]^2}\right),\nonumber
\end{align}
which agrees with the result obtained in \cite{AW}. The leading term
is the Stefan-Boltzmann term which is equal to the vacuum energy of
$\R^d$ \eqref{eq9_1}. The second term is of order $\beta^{-1}$ and
it is divergent for $p=d$. In \cite{AW}, Ambj{\o}rn and Wolfram
argued that to obtain the  $p=d$ case from this formula, one needs
to remove the divergence by subtracting the free Bose gas result,
i.e. replace the second term  by
\begin{align*}&-\frac{1}{2}\lim_{p\rightarrow d}\left(\frac{\Gamma\left(\frac{d}{2}\right)}{\pi^{\frac{d}{2}}}L_1\ldots
L_p\frac{1}{\beta}Z_{E,p}\left(\frac{d}{2}; L_1, \ldots,
L_p\right)-\frac{\Gamma\left(\frac{d-p+1}{2}\right)}{\pi^{\frac{d-p+1}{2}}}Z_{E,1}
\left(\frac{d-p+1}{2};\beta\right)\right)\\=&-\frac{1}{2\beta}\left(
Z_{E,p}'\left(0; L_1^{-1},\ldots, L_p^{-1}\right)-Z_{E,1}'(0,
\beta^{-1})\right).\end{align*} Comparing to \eqref{eq65},
 we have shown mathematically that this is indeed the case.
Using \eqref{eq4_7} and \eqref{eq9_2}, one can also obtain the high
temperature expansion of the free energy densities $f_{D/N,d}$ and
$f_{A_{C/B},d}$ for $1\leq p\leq d-1$. We find that the leading term
for all the cases is equal to the vacuum energy of $\R^d$
\eqref{eq9_1}. In the cases of Dirichlet and Neumann conditions,
there are terms proportional to $\beta^{-j}$ for every $d-p+1\leq
j\leq d+1$ as well as for $j=1$. For electromagnetic field, when $d$
is odd and $p\geq (d+1)/2$, there is no term proportional to
$\beta^{-\frac{d+1}{2}}.$

\section{Conclusion} We have provided a rigorous derivation of the Casimir free energy at finite temperature for
massless scalar fields  and electromagnetic field confined in a
closed rectangular cavity with different boundary conditions by zeta
regularization method.  By applying Chowla-Selberg formula, we
obtained explicit formulas for the low and high temperature
expansions of the free energy, which can be written as a sum of
polynomial order terms in $T$ or $T^{-1}$ plus an exponentially
decay term. To the best of our knowledge, such explicit formulas for
the low and high temperature expansions of the free energy of fields
confined within closed cavities has not been obtained previously.

We   noted that for all the cases considered, the free energy at
finite temperature $F(\beta;L_1,\ldots,L_d)$ transforms as
\begin{align*}
F(\beta;L_1,\ldots,L_d)\mapsto F(\lambda\beta;\lambda
L_1,\ldots,\lambda L_d)=\lambda^{-1}F(\beta;L_1,\ldots,L_d),
\end{align*}under the simultaneous
space--time scaling $\beta\mapsto \lambda\beta$, $L_i\mapsto \lambda
L_i, 1\leq i\leq d$. This in turn implies the thermodynamic relation
$$F=(P_1+\ldots+P_d)V-TS,$$which has not  been observed.

On the other hand, we also show that the free energy in all the
cases considered is a decreasing function of temperature.  For
massless scalar field under periodic and Neumann boundary
conditions, the free energy is negative for all temperature. For
massless scalar field under Dirichlet boundary condition and for
electromagnetic fields, the free energy might be positive at zero
temperature. When this happens, there is a unique transition
temperature at which the free energy change from positive to
negative. This transition temperature is shown graphically for $d=2$
and $d=3$. We believe that for massless scalar field under Dirichlet
boundary condition and for electromagnetic fields, when $d\geq 4$,
the zero temperature free energy will also be positive for $(L_1,
\ldots, L_d)$ lying in some domain of $\R^d$. A detail study of this
is left to another paper.

In the last section, we show how the free energy for a non-closed
rectangular cavity can be obtained by letting the size of some
directions of  a closed cavity going to infinity. We prove that the
results are in agreement with that based on the  integration
prescription usually adopted by other authors.

We remark that the discussion given in this paper focused mainly on
the low and high temperature expansions of the free energy and the
properties of the free energy. We have not dealt with other
thermodynamic quantities such as the force, pressure, internal
energy and entropy. We hope to consider these quantities in a future
work.

Finally, we would like to point out that there exist some
controversies regarding imposing boundary conditions on a quantum
field. Deutsch and Candelas \cite{DC} were the first to study the
nonintegrable divergences in the renormalized energy density near
boundaries. This problem has been re-examined  by Baacke and
Kr\"usemann \cite{BKN} and analyzed in detail recently by Jaffe
\cite{J1, J2} and Graham et al \cite{J3,J4,J5,J6}. These authors
showed that the imposition of boundary conditions on quantum fields
in Casimir effect calculations leads to non-renormalizable
infinities. As a result, fixing boundary conditions \emph{ab initio}
invariably results in divergences which cannot be removed by
renormalization. Basically this problem for electromagnetic field
with Dirichlet boundary condition can be stated as that no real
material is perfectly conducting at arbitrary high frequencies. In
order to overcome this serious problem, Graham and collaborators
have developed a new approach which replaces the boundary condition
by a renormalizable coupling between the fluctuating field and a
non-dynamical background field representing the material. On the
other hand, there were responses from Milton \cite{Mil}, Fulling
\cite{Fu} and Elizalde \cite{Ee} with various attempts to resolve
this issue. Here we would like to mention the effort by Elizalde who
has tried to explain the presence of infinities as a result of
drastic reduction of eigenstates when boundary condition is imposed.
He has proposed to complement the zeta function method with the
Hadamard regularization in order to make sense of infinities present
in the boundary value problems in Casimir energy calculations.
However such an approach cannot be taken as a substitute of the more
physical treatment given in ref. \cite{J1,J2,J3,J4,J5,J6}. The
system considered in this paper can be regarded as ideal cases, for
which zeta function technique is still a useful tool for
regularization of vacuum energy density. For a more physical
treatment, one has no choice but have to take into account of the
problem of singular behavior near a boundary.

\vspace{0.3cm} \noindent \textbf{Acknowledgement}\; The authors
would like to thank Malaysian Academy of Sciences, Ministry of
Science, Technology  and Innovation for funding this project under
the Scientific Advancement Fund Allocation (SAGA) Ref. No P96c.

\appendix
\section{}In this appendix, we gather some mathematical formulas and
estimates that we need.

\vspace{0.2cm} \noindent 1. We want to prove \eqref{eq27_3}. By
equation \eqref{eq3_1}, we find that
\begin{align*}
F_{A_B}\left(L_1,\ldots, L_d\right)=&\sum_{j=1}^{d}
c_{j;d}\sum_{1\leq m_1<\ldots<m_j\leq d}F_P\left(L_{m_1}, \ldots,
L_{m_j}\right),
\end{align*}
\begin{align*}
\text{where}\hspace{1cm}c_{j;d}=
\sum_{k=j}^{d}(-1)^{k-j}(k-1)\begin{pmatrix}
d-j\\k-j\end{pmatrix}2^{-k}.
\end{align*}Now we compute $c_{j;d}$.
\begin{align*}
c_{j;d}=& \sum_{k=0}^{d-j} (-1)^k (k+j-1)\begin{pmatrix} d-j\\k
\end{pmatrix}2^{-k-j}\\
=&2^{-j}\left(\sum_{k=0}^{d-j} (-1)^k k\begin{pmatrix} d-j\\k
\end{pmatrix}2^{-k}+(j-1)\sum_{k=0}^{d-j} (-1)^k \begin{pmatrix} d-j\\k
\end{pmatrix}2^{-k}\right)\\
=&2^{-j}\left((d-j)\sum_{k=1}^{d-j} (-1)^k \begin{pmatrix}
d-j-1\\k-1
\end{pmatrix}2^{-k}+(j-1)\sum_{k=0}^{d-j} (-1)^k \begin{pmatrix} d-j\\k
\end{pmatrix}2^{-k}\right)\\
=&2^{-j}\left(-\frac{(d-j)}{2}\sum_{k=0}^{d-j-1} (-1)^k
\begin{pmatrix} d-j-1\\k
\end{pmatrix}2^{-k}+(j-1)\sum_{k=0}^{d-j} (-1)^k \begin{pmatrix} d-j\\k
\end{pmatrix}2^{-k}\right)\\
=&2^{-j}\left(-\frac{(d-j)}{2}\left(1-\frac{1}{2}\right)^{d-j-1}+(j-1)\left(1-\frac{1}{2}\right)^{d-j}\right)\\
=&2^{-d}(2j-d-1).
\end{align*}

\vspace{0.2cm} \noindent 2. We want to show that
\begin{align*}
\lim_{\substack{a_i\rightarrow 0\\1\leq i\leq m
}}\left[\prod_{j=1}^ma_j\right]R_{n,m}(a_1, \ldots,a_n)=0,
\end{align*}with $R_{n,m}$ defined by \eqref{eq30_4}. Without loss of generality, we assume that $a_1\leq \ldots\leq a_n$. Define
$\alpha_1(\mathbf{k})=
\sqrt{\sum_{j=1}^m\left[\frac{k_j}{a_j}\right]^2}$,
$\alpha_2(\mathbf{k})=\sqrt{\sum_{j=m+1}^n[a_{j}k_{j}]^2}$. Then by
\eqref{eq30_4} and using \begin{align}\label{eq30_6}
|K_{\nu}(z)|\leq &
\sqrt{\frac{\pi}{2z}}e^{-z}\Biggl[1+\sum_{k=1}^{[\nu]}\frac{1}{(2z)^k
k!}\prod_{j=1}^{k}\left(\nu^2-\left[\frac{2j-1}{2}\right]^2\right)\Biggr]\\
\leq &
\sqrt{\frac{\pi}{2z}}e^{-z}\sum_{k=0}^{[\nu]}\frac{c_{\nu}}{(2z)^k
k!},\nonumber
\end{align}where $$c_{\nu}=4\prod_{j=1}^{[\nu]}\left(\nu^2-\left[\frac{2j-1}{2}\right]^2\right),$$ we have
\begin{align*}
\left|\tilde{R}_{n,m}\right|=&\left|\left[\prod_{j=1}^ma_j\right]R_{n,m}(
a_1, \ldots, a_n)\right|\\
\leq &c_{ \frac{m}{2} }\sum_{\mathbf{k}\in(\Z^m\setminus
\{\mathbf{0}\})\times(\Z^{n-m}\setminus\{\mathbf{0}\})}e^{-2\pi\alpha_1(\mathbf{k})\alpha_2(\mathbf{k})}
\sum_{l=0}^{\left[\frac{m}{2}\right]}\frac{1}{(4\pi)^ll!}\alpha_1(\mathbf{k})^{-l
-\frac{m+1}{2}}\alpha_2(\mathbf{k})^{-l+\frac{m-1}{2}}.\\
\end{align*}Using the inequality
$$\sqrt{\sum_{j=1}^{n} x_j^2}\geq \frac{1}{\sqrt{n}}\left(\sum_{j=1}^{n} |x_j|\right)\geq \sqrt{n}\min\{|x_j|\},$$we have
\begin{align*}
&\sum_{\mathbf{k}\in(\Z^m\setminus
\{\mathbf{0}\})\times(\Z^{n-m}\setminus\{\mathbf{0}\})}e^{-2\pi\alpha_1(\mathbf{k})\alpha_2(\mathbf{k})}
\alpha_1(\mathbf{k})^{-l
-\frac{m+1}{2}}\alpha_2(\mathbf{k})^{-l+\frac{m-1}{2}}\\\leq &
\sum_{\mathbf{k}\in(\Z^m\setminus
\{\mathbf{0}\})\times(\Z^{n-m}\setminus\{\mathbf{0}\})}e^{-\frac{2\pi\alpha_2(\mathbf{k})}{\sqrt{m}}\sum_{j=1}^{m}\left|
\frac{k_j}{a_j}\right|}a_m^{l+\frac{m+1}{2}}
\alpha_2(\mathbf{k})^{-l+\frac{m-1}{2}}\\
=&a_m^{l+\frac{m+1}{2}}\sum_{(k_{m+1}, \ldots,
k_n)\in\Z^{n-m}\setminus\{\mathbf{0}\}}\alpha_2(\mathbf{k})^{-l+\frac{m-1}{2}}\left(
\left[1+\frac{2e^{-\frac{2\pi\alpha_2(\mathbf{k})}{a_m\sqrt{m}}}}
{1-e^{-\frac{2\pi\alpha_2(\mathbf{k})}{a_m\sqrt{m}}}}\right]^{m}-1\right)\\
\leq &2ma_m^{l+\frac{m+1}{2}}\frac{\left(1+e^{-\frac{2\pi
a_{m+1}}{a_m\sqrt{m}}}\right)^{m-1}}{\left(1-e^{-\frac{2\pi
a_{m+1}}{a_m\sqrt{m}}}\right)^{m}}\sum_{(k_{m+1}, \ldots,
k_n)\in\Z^{n-m}\setminus\{\mathbf{0}\}}\alpha_2(\mathbf{k})^{-l+\frac{m-1}{2}}e^{-\frac{2\pi\alpha_2(\mathbf{k})}{a_m\sqrt{m}}}.
\end{align*}From this, it is easily seen that as $a_i\rightarrow 0$
for $1\leq a_i\leq m$, $\tilde{R}_{n,m}\rightarrow 0$.

\vspace{0.2cm} \noindent 3. We want to show that the integral
\begin{align*}
I=\frac{1}{(2\pi)^{d-p}\beta}\int_{\R^{d-p}}\sum_{\mathbf{k}\in
\Z^{p}}\log\left(1-e^{-\beta\sqrt{\sum_{j=1}^p\left[\frac{2\pi
k_j}{L_j}\right]^2+ |\mathbf{w}|^2}}\right)dw_1\ldots
dw_{d-p}\end{align*} is equal to
\begin{align}\label{eq8_1}
&-\frac{\Gamma\left(\frac{d-p+1}{2}\right)}{\pi^{\frac{d-p+1}{2}}}\frac{\zeta_R(d-p+1)
}{\beta^{d-p+1}} \\
&-\frac{2}{\beta^{\frac{d-p+1}{2}}}\sum_{m=1}^{\infty}\sum_{\mathbf{k}\in
\Z^{p}\setminus\{\mathbf{0}\}}\frac{1}{m^{\frac{d-p+1}{2}}}\left(\sum_{j=1}^p
\left[\frac{k_j}{L_j}\right]^2\right)^{\frac{d+1-p}{4}}K_{\frac{d-p+1}{2}}\left(2\pi
m\beta \sqrt{\sum_{j=1}^p
\left[\frac{k_j}{L_j}\right]^2}\right).\nonumber
\end{align}We split $I$ into two terms $I_1$ and $I_2$, where $I_1$
corresponds to $\mathbf{k}=\mathbf{0}$ term and $I_2$ contains the
$\mathbf{k}\in \mathbb{Z}^p\setminus\{\mathbf{0}\}$ terms. We have
\begin{align*}
I_1=&\frac{1}{(2\pi)^{d-p}\beta}\int_{\R^{d-p}}\log\left(1-e^{-\beta|\mathbf{w}|}\right)dw_1\ldots
dw_{d-p}\\
=&-\frac{2\pi^{\frac{d-p}{2}}}{\Gamma\left(\frac{d-p}{2}\right)(2\pi)^{d-p}\beta}
\int_0^{\infty} w^{d-p-1} \sum_{m=1}^{\infty}\frac{e^{-m\beta
w}}{m}dw\\
=&-\frac{\Gamma\left(d-p\right)}{\Gamma\left(\frac{d-p}{2}\right)2^{d-p-1}\pi^{\frac{d-p}{2}}\beta^{d-p+1}}\sum_{m=1}^{\infty}
\frac{1}{m^{d-p+1}}.
\end{align*}Using the formula
$\Gamma(2z)=2^{2z-1}\pi^{-1/2}\Gamma(z)\Gamma\left(z+\frac{1}{2}\right)$
(8.335 of \cite{GR}), we find that $I_1$ is equal to
\begin{align*}
I_1=-\frac{\Gamma\left(\frac{d-p+1}{2}\right)}{\pi^{\frac{d-p+1}{2}}\beta^{d-p+1}}\zeta_R(d-p+1).
\end{align*}For $I_2$, set $v(\mathbf{k})=\sqrt{\sum_{j=1}^p\left[\frac{2\pi
k_j}{L_j}\right]^2}$, we have
\begin{align*}
I_2=&\frac{1}{(2\pi)^{d-p}\beta}\int_{\R^{d-p}}\sum_{\mathbf{k}\in
\Z^{p}\setminus\{\mathbf{0}\}}\log\left(1-e^{-\beta\sqrt{v(\mathbf{k})^2+
|\mathbf{w}|^2}}\right)dw_1\ldots dw_{d-p}\\
=&-\frac{2\pi^{\frac{d-p}{2}}}{\Gamma\left(\frac{d-p}{2}\right)(2\pi)^{d-p}\beta}\sum_{m=1}^{\infty}\frac{1}{m}\sum_{\mathbf{k}\in
\Z^{p}\setminus\{\mathbf{0}\}}\int_{0}^{\infty}w^{d-p-1}e^{-m\beta\sqrt{v(\mathbf{k})^2+w^2}}dw.
\end{align*}Now using the substitution $u=\sqrt{v^2+w^2}$ and the formula 4 of 3.389 in \cite{GR}, we have
\begin{align*}
\int_{0}^{\infty}w^{d-p-1}e^{-m\beta\sqrt{v^2+w^2}}dw=&\int_v^{\infty}u
(u^2-v^2)^{\frac{d-p}{2}-1}e^{-m\beta u} du\\
=&2^{\frac{d-p-1}{2}}\pi^{-\frac{1}{2}}\frac{1}{(m\beta)^{\frac{d-p-1}{2}}}v^{\frac{d-p+1}{2}}\Gamma\left(
\frac{d-p}{2}\right)K_{\frac{d-p+1}{2}}\left(m\beta v\right).
\end{align*}Combining together we find that $I_2$ is equal to the
second term in \eqref{eq8_1}, thus proving our claim.

\end{document}